\PassOptionsToPackage{protrusion, expansion}{microtype} 

\documentclass[sigconf,screen]{acmart}

\usepackage{amsmath, amsthm}
    
\usepackage{graphicx, xcolor} 
\usepackage{hyperref, url}
\usepackage{adjustbox}
\usepackage{algorithmic}
\usepackage{algorithm}
\usepackage{comment}
\usepackage{booktabs, multirow}
\usepackage{caption}
\usepackage[tight,footnotesize]{subfigure}
\usepackage{enumitem}
\usepackage{lipsum}
\usepackage{setspace}
\usepackage{diagbox}
\usepackage{soul}
\usepackage{textcomp}
\usepackage{xspace}
\usepackage{pifont}
\usepackage[many]{tcolorbox}

\usepackage{tikz}

\usepackage{footnote}
\makesavenoteenv{tabular}
\makesavenoteenv{table}

\usepackage{graphicx}

\usepackage[acronym]{glossaries}
\makeglossaries
 
\newglossaryentry{fclayer}{
    name={\texttt{fc}-layer},
    description={fully connected layer}
}
 
\newglossaryentry{szfull}{
    name=\textsc{SZ lossy compression},
    description={SZ lossy compression}
}

\newglossaryentry{deepsz}{
    name=\textsc{DeepSZ},
    description={DeepSZ}
}

\newglossaryentry{dataArray}{
    name=\textsc{data array},
    description={}
}

\newglossaryentry{indexArray}{
    name=\textsc{index array},
    description={}
}
 
\newacronym{gcd}{GCD}{Greatest Common Divisor}
\newacronym{lcm}{LCM}{Least Common Multiple}
 
\newcommand{\tech}{\textsc{PaRID}\xspace}

\graphicspath{{./figures/}}
\DeclareGraphicsExtensions{.pdf}

\setcopyright{none}
\copyrightyear{2022}
\acmYear{2022}

\definecolor{main}{HTML}{464646}    
\definecolor{sub}{HTML}{efefef}     

\definecolor{main1}{HTML}{464646}    
\definecolor{sub1}{HTML}{efefef}     

\tcbset{
    sharp corners,
    colback = white,
    before skip = 0.2cm,    
    after skip = 0.5cm      
}                           

\newtcolorbox{boxA}{
    fontupper = \bf,
    boxrule = 1.5pt,
    colframe = black 
}

\newtcolorbox{boxB}{
    fontupper = \bf\color{main}, 
    boxrule = 1.5pt,
    colframe = main,
    rounded corners,
    arc = 5pt   
}

\newtcolorbox{boxC}{
    colback = sub, 
    boxrule = 0pt  
}

\newtcolorbox{boxD}{
    colback = sub, 
    colframe = main, 
    boxrule = 0pt, 
    toprule = 3pt, 
    bottomrule = 3pt 
}

\newtcolorbox{boxE}{
    enhanced, 
    boxrule = 0pt, 
    borderline = {0.75pt}{0pt}{main}, 
    borderline = {0.75pt}{2pt}{sub} 
}

\newtcolorbox{boxF}{
    colback = sub,
    enhanced,
    boxrule = 1.5pt, 
    colframe = white, 
    borderline = {1.5pt}{0pt}{main, dashed} 
}

\newtcolorbox{boxG}{
    enhanced,
    boxrule = 0pt,
    colback = sub,
    borderline west = {1pt}{0pt}{main}, 
    borderline west = {0.75pt}{2pt}{main}, 
    borderline east = {1pt}{0pt}{main}, 
    borderline east = {0.75pt}{2pt}{main}
}

\newtcolorbox{boxH}{
    colback = sub, 
    colframe = main, 
    boxrule = 0pt, 
    leftrule = 4pt 
}

\newtcolorbox{boxHH}{
    colback = sub1, 
    boxrule = 0pt  
}

\newtcolorbox{boxI}{
    colback = sub, 
    colframe = main, 
    boxrule = 0pt, 
    toprule = 6pt 
}

\newtcolorbox{boxJ}{
    sharpish corners, 
    colback = sub, 
    colframe = main, 
    boxrule = 0pt, 
    toprule = 4.5pt, 
    enhanced,
    fuzzy shadow = {0pt}{-2pt}{-0.5pt}{0.5pt}{black!35} 
}

\newtcolorbox{boxK}{
    sharpish corners, 
    boxrule = 0pt,
    toprule = 4.5pt, 
    enhanced,
    fuzzy shadow = {0pt}{-2pt}{-0.5pt}{0.5pt}{black!35} 
}

\newtcolorbox{boxL}{
    fontupper = \color{main},
    rounded corners,
    arc = 6pt,
    colback = sub, 
    colframe = main!50, 
    boxrule = 0pt, 
    bottomrule = 4.5pt 
}

\newtcolorbox{boxM}{
    fontupper = \color{white},
    rounded corners,
    arc = 6pt,
    colback = main!80, 
    colframe = main, 
    boxrule = 0pt, 
    bottomrule = 4.5pt,
    enhanced,
    fuzzy shadow = {0pt}{-3pt}{-0.5pt}{0.5pt}{black!35}
}

\begin{document}
\title{Detecting Soft Errors in Parallel Software with LLM-tuned Instruction Duplication}

\author{Yafan Huang}
\affiliation{%
  \institution{University of Iowa}
  \city{Iowa City}
  \state{IA}
  \country{USA}
}
\email{yafan-huang@uiowa.edu}

\author{Guanpeng Li}
\affiliation{%
  \institution{University of Florida}
  \city{Gainesville}
  \state{FL}
  \country{USA}
}
\email{liguanpeng@ufl.edu}

\begin{abstract}
We propose \tech (PaRallel Instruction Duplication), a software-directed soft error detection framework that requires only compile-time efforts for multithreading parallel programs. \tech addresses two key challenges: supporting parallel programs with mixed serial and parallel regions and minimizing performance overhead without relying on costly dynamic profiling. It combines parallel-aware code transformation with LLM-tuned performance modeling, guided by eight generalizable findings from an offline characterization study, to enable fast soft error detection in parallel applications. Evaluation on NPB benchmarks shows that \tech reduces protection overhead from 162.79\% to 59.84\% on average and achieves up to 5$\times$ speedup while maintaining full error detection effectiveness.
\end{abstract}

\keywords{Software-directed Reliability, Soft Error Detection, Compiler Code Transformation, Parallel Programs, Large Language Models}

\maketitle

\setlength{\textfloatsep}{6pt}

\section{Introduction}

With the increasing scale of computing infrastructures, error resilience is becoming a critical concern in not only modern high-performance computing systems~\cite{guo2018fliptracker,lucas2014doe}, but also industrial data centers operated by companies such as Google~\cite{hochschild2021cores} and Meta~\cite{dixit2021silent}.
A recent report from Frontier~\cite{atchley2023frontier}, the first exascale system deployed at Oak Ridge National Laboratory, highlights this growing challenge: despite its advancements in energy efficiency, memory hierarchy, and massive concurrency, Frontier yet exhibits 2–3$\times$ lower error resilience compared to early terascale systems~\cite{bergman2008exascale}.
Among the various sources of faults, soft errors, also known as transient hardware faults, are particularly insidious~\cite{shivakumar2002modeling,agiakatsikas2023impact}.
These errors often originate as an unexpected bit-flip in hardware components and can silently propagate through the system stack, ultimately corrupting application outputs without any observable warning~\cite{dixit2021silent}.


In many real‑world applications, producing incorrect computation results can have catastrophic consequences.
In smoothed particle hydrodynamics simulation~\cite{monaghan2005smoothed}, which is a numerical method used to predict extreme flood events, a single bit-flip soft error can propagate through interpolation to $\sim$100 particles~\cite{cavelan2019detection}, leading to incorrect predictions of flood impact zones, exposing communities to life-threatening risks~\cite{mirauda2020smoothed}.
As such, soft error detection becomes an essential need.
Although various detection techniques have been proposed, many suffer from notable drawbacks. 
Hardware‑based schemes impose high energy costs~\cite{kim2010soft}, while algorithm‑specific checks are tailored to certain computation patterns (e.g., matrix multiplication)~\cite{huang1984algorithm}, limiting their general applicability.
In contrast, \textit{instruction duplication}, a software-directed approach, has emerged as a promising solution~\cite{oh2002error,reis2005swift,wei2014quantifying,laguna2016ipas,kalra2020armorall,mahmoud2018optimizing}.
By duplicating instructions at compile time and checking value mismatches at runtime, this technique can effectively detect soft errors in a lightweight, algorithm-agnostic manner.

\begin{figure}[h]
	\centering
    \hspace{1mm}
	\subfigure
	{\includegraphics[width=0.282\columnwidth]{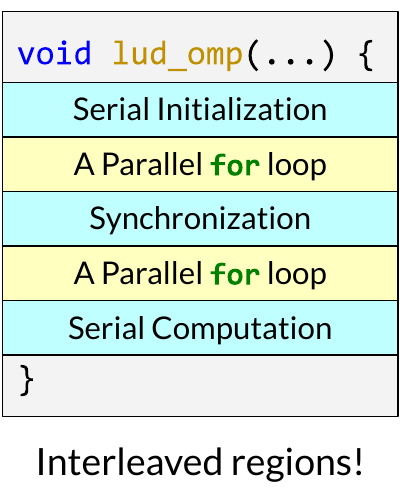}\label{fig:challenge-1-mixed-region}}
    \subfigure
	{\includegraphics[width=0.32\columnwidth]{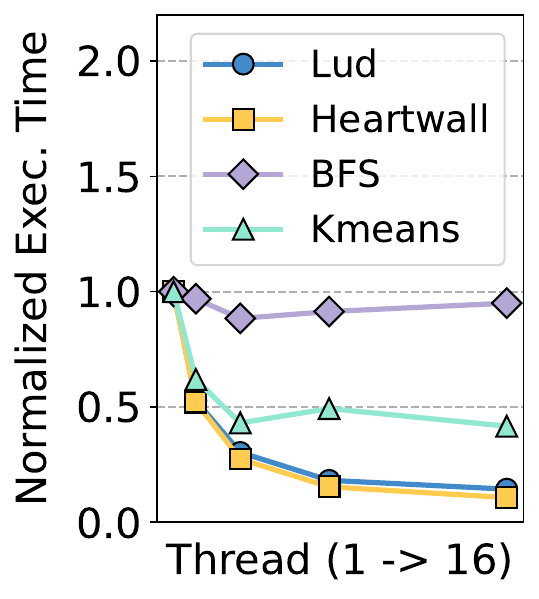}\label{fig:challenge-2-1}}
    \subfigure
	{\includegraphics[width=0.32\columnwidth]{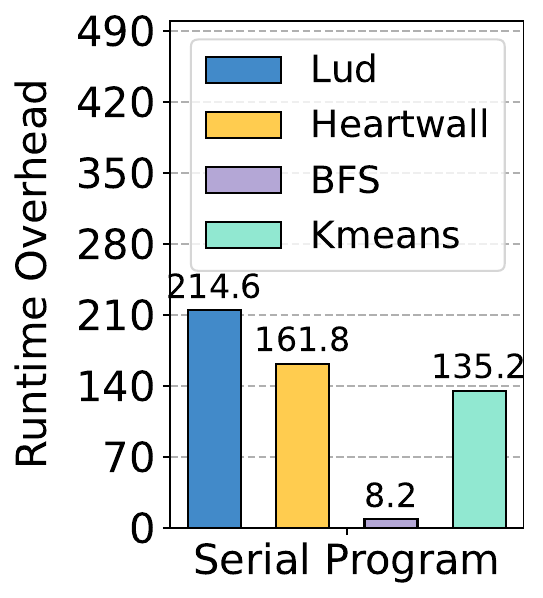}\label{fig:challenge-2-2}}
    \vspace{-4mm}
    \caption{Two challenges for performing efficient instruction duplications for multithreading parallel programs. \textbf{C-1} (left): interleaved serial and parallel regions; \textbf{C-2} (middle): thread count impacts differently on runtime performance across programs; \textbf{C-2} (right): existing serial instruction duplication impacts differently on runtime overhead for programs.}
    \label{fig:intro-challenges}
\end{figure}

In the past, instruction duplication has been studied across various scenarios, spanning CPU and GPU architectures~\cite{reis2005swift,kuvaiskii2016elzar,kalra2020armorall,mahmoud2018optimizing,yim2011hauberk,didehban2023generic} and covering both selective and full duplication~\cite{chen2016simd,huang2023characterizing,laguna2016ipas,lu2014sdctune}.
However, a critical gap remains before instruction duplication can be widely adopted in real-world settings: its applicability in \textit{multithreading parallel CPU environments}.
This setting is vital in accelerating many workloads, such as stencil computations in seismic imaging~\cite{raut2020evaluating}.
Achieving effective and efficient instruction duplication for parallel programs is non-trivial and requires addressing two fundamental challenges.
\textbf{C-1: Mixed Serial and Parallel Regions.}
Unlike GPU programs, which are fully parallelized within GPU kernels, or traditional serial CPU programs, multithreading parallel CPU applications often consist of a mix of serial and parallel regions that may be nested or interleaved during execution~\cite{dagum1998openmp,de2018ongoing}.
Figure~\ref{fig:challenge-1-mixed-region} illustrates this with the core function \verb|lud_omp()| for Lud program from Rodinia~\cite{che2009rodinia}.
As seen, parallel regions, serial regions, and synchronization points are interleaved across this function.
An effective duplication strategy must correctly handle this hybrid structure by ensuring thread-safe duplication and managing inter-thread orchestrations.
To date, none of the existing duplication techniques thoroughly address these requirements~\cite{huang2023characterizing,he2023demystifying,laguna2016ipas,lu2014sdctune,didehban2023generic}.
\textbf{C-2: Minimizing Performance Loss with Detection.}
In parallel software, efficiency is often the top priority--parallelism is introduced explicitly to accelerate execution~\cite{skinner2005performance}.
However, combining instruction duplication with parallelism introduces significant uncertainty in runtime performance.
First, the sensitivity of performance to thread count varies widely across programs; a configuration that performs well for one may be suboptimal for another.
In Figure~\ref{fig:challenge-2-1}, increasing the thread count from 4 to 16 yields significant speedups for Lud, Heartwall, and Kmeans, but results in slower execution for BFS.
Second, even within the same program, different parallel regions may react differently to thread count—some scale efficiently, while others degrade quickly~\cite{furlinger2007scalability}.
Additionally, the overhead introduced by instruction duplication is highly program-specific, ranging from 8.25\% in BFS to over 200\% in LU (see Figure~\ref{fig:challenge-2-2}), and becomes even more elusive in parallel programs.
Existing approaches often rely on trial-and-error to identify optimal thread settings~\cite{marathe2015run}.
However, profiling such information demands substantial real-time effort, an impractical burden in production, where workloads routinely run from hours to months~\cite{shaw2021anton}.
Repeated executions with different thread counts can be prohibitively expensive, making traditional tuning methods infeasible for large-scale applications.

To tackle these challenges, we propose \tech (\textit{\textbf{PaR}allel \textbf{I}nstruction \textbf{D}uplication}).
Given an arbitrary parallel program code, \tech uses two independent steps to generate the duplicated executable binary.
While \textit{Parallel-aware Code Transformation} instruments duplicated and other functional instructions safely for parallel programs, \textit{LLM-tuned Performance Modeling} automatically locates the best thread settings, no matter how large the workload is, with prompt-refined large language model (LLM) inferences, both of which require only compile-time efforts without dynamic profiling.
\textit{To the best of our knowledge, this is not only the first work to comprehensively propose an effective instruction duplication for parallel programs but also the first work to predict dynamic program features with LLM in software error resilience.}
Some key results of \tech can be found as follows.
\begin{itemize}[leftmargin=8mm]
    \item \tech’s Parallel-Aware Code Transformation exhibits high compatibility across all 15 Rodinia benchmarks and 8 NPB benchmarks, under diverse thread counts and input sizes.
    \item We summarize 8 \textit{Findings} from an offline thread-sensitivity study to guide LLMs in selecting optimal thread settings. These findings are generalizable across different machines.
    \item Without thread tuning from LLM, instruction duplication for parallel programs incurs an average execution time of \textbf{162.79\%} over serial execution. \tech reduces this to \textbf{59.84\%}, achieving up to $\sim$5$\times$ speedup (on NPB EP).
    \item All benefits require only compile-time efforts and do not sacrifice error detection effectiveness.
\end{itemize}

\section{Background}



\subsection{Fault Model and Compiler Platform} 
\label{sec:fault-model}
In this work, we focus on soft errors that occur in the processor's computational pipeline (i.e., datapath units), including arithmetic logic units, pipeline stages, and load/store units.
We exclude errors in memory and cache units from our scope, as they can be effectively mitigated by techniques such as error-correcting codes (ECC)~\cite{chou2015reducing,alam2022comet,tsai2012study,patel2020bit} and parity checks~\cite{plank2004practical}.
Similarly, we exclude errors that corrupt control-flow units (e.g. program counter) since software signatures~\cite{oh2002control,goloubeva2003soft} can handle them.
Soft errors manifesting at the program level are considered single bit-flips, as the occurrence rate of multiple bit-flips is relatively low in existing systems~\cite{bautista2016unprotected}.
Our fault model is consistent with existing works that study soft error detection mechanisms~\cite{oh2002error,reis2005swift,reis2007automatic,laguna2016ipas,li2018modeling,kalra2020armorall}. 

We utilize LLVM compiler~\cite{lattner2004llvm} for performing instruction duplication and fault injection due to its standard optimizer and support for multiple programming languages and hardware platforms.
LLVM features a human-readable intermediate representation (IR) with instruction-like characteristics, enabling users to perform various transformations.
Additionally, LLVM supports optimizations for OpenMP~\cite{dagum1998openmp} starting from v11~\cite{llvm-openmp}.
Using LLVM compiler and IR to explore program error resilience has been widely adopted in existing reliability studies~\cite{yang2021enabling,kuvaiskii2016elzar,laguna2016ipas,spensky2021glitching,kalra2020armorall,mahmoud2018optimizing,guan2020chaser,huang2022mitigating}.

\subsection{The Principle of Instruction Duplication}
\begin{figure}[h]
\centering
\includegraphics[width=1.0\linewidth]{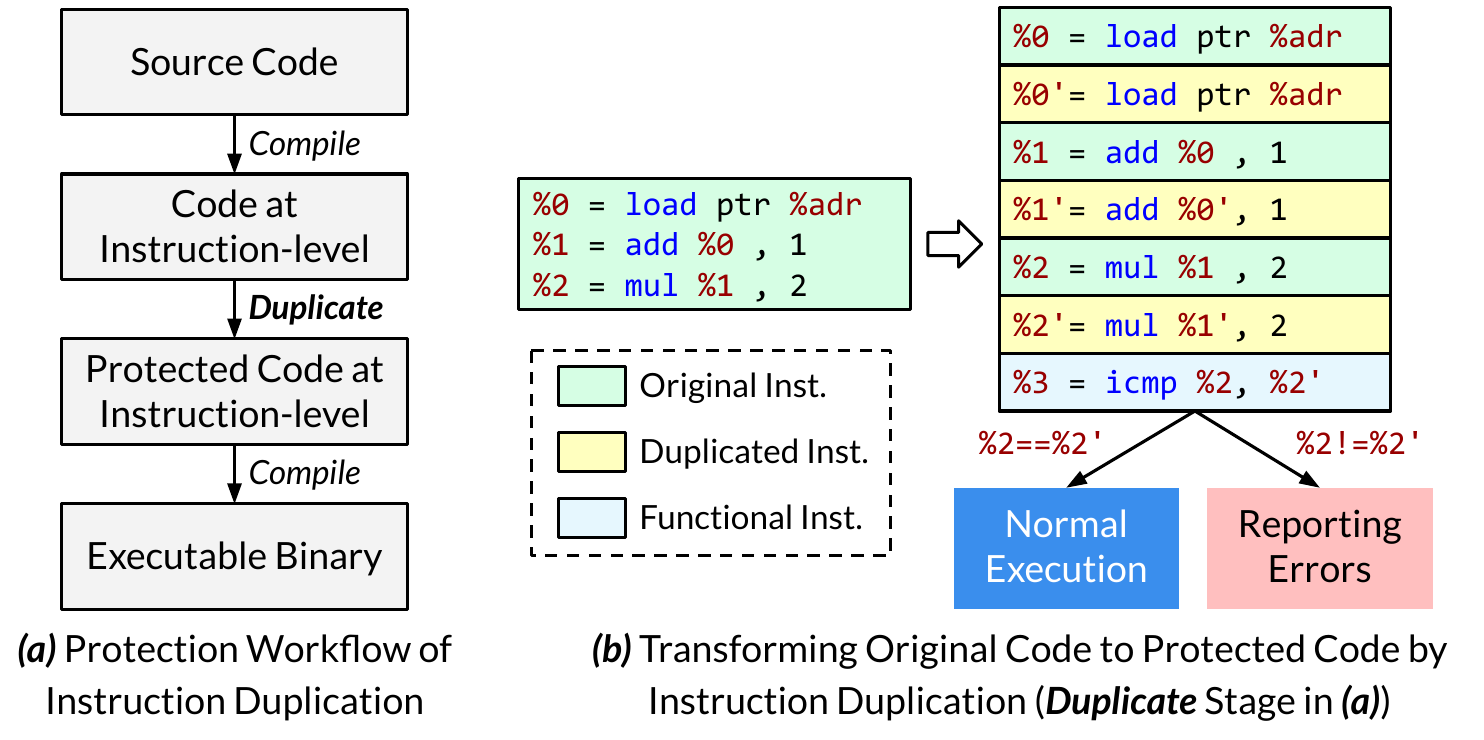}
\vspace{-6mm}
\caption{Illustration of instruction duplication.}
\label{fig:instruction-duplication}
\end{figure}

Instruction duplication is a soft error detection technique that requires only compiler-level program-agnostic code transformation.
Figure~\ref{fig:instruction-duplication}(a) illustrates the protection workflow of instruction duplication.
After the compiler compiles the source code into an instruction format, instruction duplication instruments replicas, generating protected code. This protected code is finally compiled into an executable binary for deployment.
Instruction duplication is a flexible software-level solution that requires only static code transformation at the compiler level, ensuring this technique is program-agnostic and does not require expensive dynamic profiling.
As a result, instruction duplication has been actively studied over the past two decades~\cite{oh2002error,reis2005swift,lu2014sdctune,laguna2016ipas,mahmoud2018optimizing,kalra2020armorall} and successfully applied in real-world safety-critical missions, such as the fault-tolerant systems used in Stanford's Advanced Research and Global Observation Satellite (ARGOS) project~\cite{lovellette2002strategies}.

Figure~\ref{fig:instruction-duplication}(b) further details a code example to explain the principle of instruction duplication.
Given a basic block\footnote{Basic block is a compiler concept and consists of multiple sequentially executed instructions. Basic blocks form the vertices in control-flow graph.} with three instructions, instruction duplication inserts replicas that use different registers (e.g. {\small\texttt{\%0}} to {\small\texttt{\%0'}}), forming a separate but functionally identical dataflow. 
A comparison instruction (e.g. {\small\texttt{icmp}}) is added at the end of the two dataflows to verify consistency. If the values match, execution proceeds to the next basic block with normal execution. Otherwise, an error is reported, triggering recovery mechanisms such as checkpoint/restart~\cite{hursey2007design}.
In this work, we target full protection -- duplicating all eligible instructions -- whereas our approach can be easily extended to support selective protection~\cite{lu2014sdctune,laguna2016ipas}.

\section{Initial Study}

The goal is to propose an \textbf{effective and efficient instruction duplication technique for parallel software}, with only compile-time effort.
To ensure effectiveness, instructions must be safely duplicated across mixed serial and parallel regions, addressing \textbf{C-1}.
To ensure efficiency, the system must provide thread configurations with low runtime overhead, not only from the program itself but also from the duplication, addressing \textbf{C-2}.
Importantly, the proposed method should avoid reliance on dynamic profiling (i.e., features that require program execution), which is often impractical in production where applications may run for extended periods~\cite{cavelan2019detection,raut2020evaluating,huang2022mitigating}.
In this section, we conduct an initial study to better understand these challenges; the key insights derived from this study directly inform the design of our proposed solution.

\begin{figure}[h]
\centering
\includegraphics[width=1.0\linewidth]{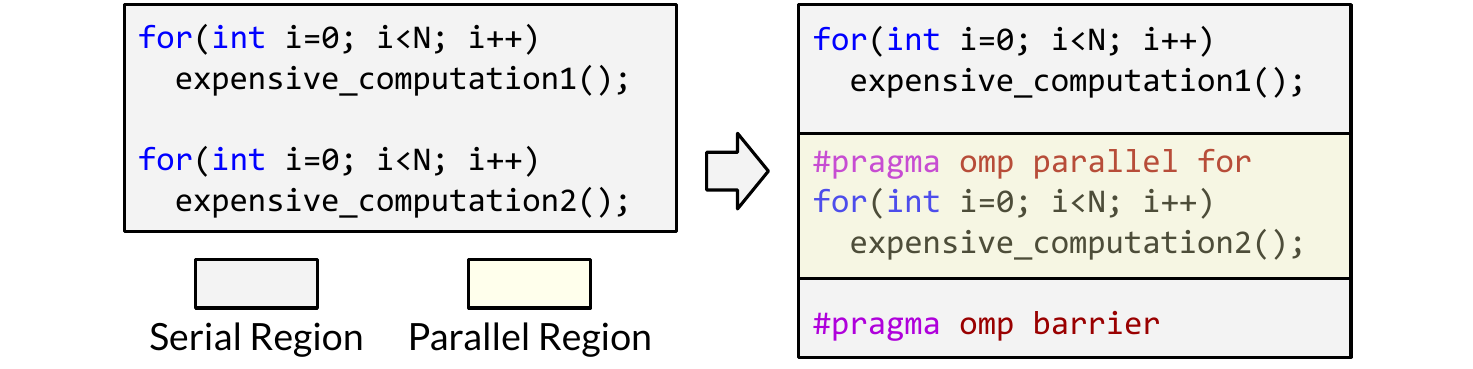}
\vspace{-5mm}
\caption{Parallel a \texttt{for} loop using OpenMP directive.}
\label{fig:using-openmp}
\end{figure}

In this work, we adopt OpenMP~\cite{dagum1998openmp} as the parallel interface. 
OpenMP is an annotation-based parallelization API for multiple programming languages, including C/C++ and Fortran, making it popular for parallel computing tasks in both academia and industry~\cite{ayguade2008design,lehromptest}.
A recent work~\cite{kadosh2023quantifying} also reports that OpenMP occupies 45\% of recent parallel analyses.
Figure~\ref{fig:using-openmp} shows an example of paralleling a loop using an OpenMP directive {\small\texttt{\#pragma omp parallel}}.

\subsection{Static Analysis of OpenMP Programs}
\label{sec:understanding}

\begin{figure}[h]
\centering
\includegraphics[width=1.0\linewidth]{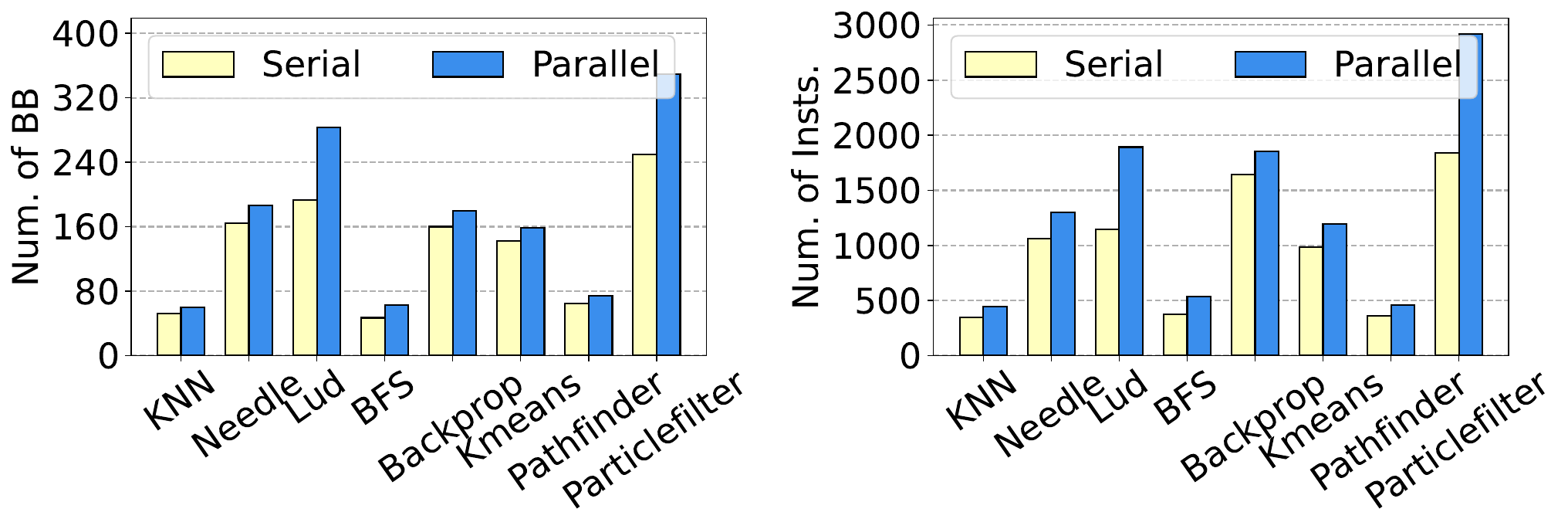}
\vspace{-8mm}
\caption{The number of static basic blocks (left) and instructions (right) for 8 Rodinia serial/parallel programs~\cite{che2009rodinia}.}
\label{fig:static-analysis}
\end{figure}

Unlike serial code, OpenMP programs consist of interleaved parallel and serial regions, making existing instruction duplication techniques inadequate~\cite{lu2014sdctune,huang2022mitigating}.
We demystify such discrepancy by performing static analysis on eight programs from the Rodinia benchmark suite~\cite{che2009rodinia}.
Specifically, we compare the number of basic blocks and instructions in serial and OpenMP-parallelized versions, compiled using the same LLVM toolchain.
As shown in Figure~\ref{fig:static-analysis}, the parallel versions consistently introduce more complexity.
On average, basic block count increases by 26\%, from 134.13 to 169.25, with a similar trend observed in instruction count (from 970.35 to 1324.38).
These additional instructions often involve parallel constructs, such as synchronization, data sharing, and thread management~\cite{quinn1994parallel}, constraining existing instruction duplication in parallel programs.

\begin{figure}[h]
\centering
\hspace{2.5mm}
\includegraphics[width=0.95\linewidth]{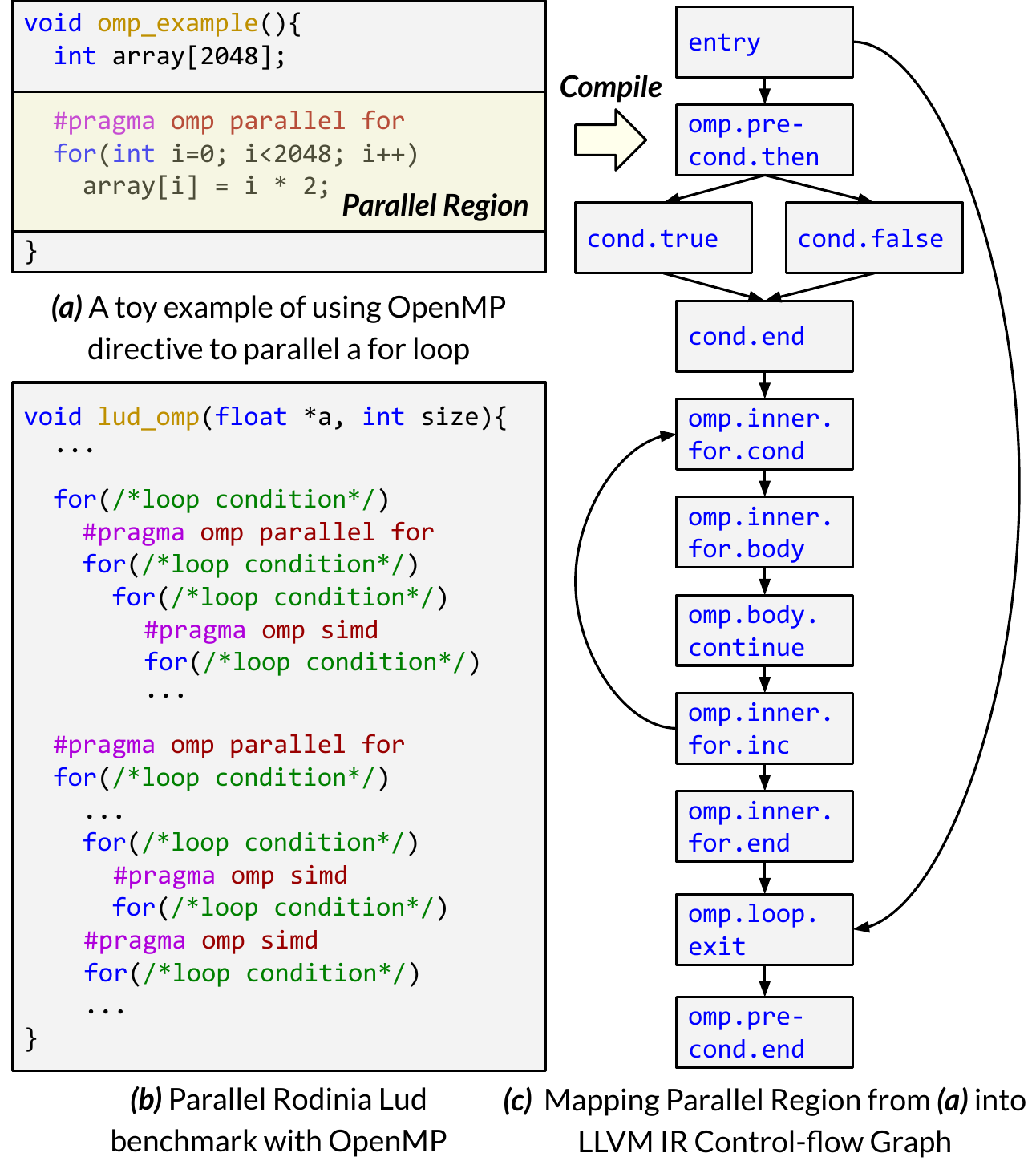}
\vspace{-2mm}
\caption{Illustration of paralleling programs with OpenMP directives and how OpenMP parallel regions reflect at LLVM.}
\label{fig:openmp-llvm}
\end{figure}

To further understand how OpenMP parallelization leads to increased instruction-level complexity, we begin with a simple {\small\texttt{for}} loop parallelized using OpenMP and examine its corresponding instruction-level representation.
The source code example is shown in Figure~\ref{fig:openmp-llvm}(a), and its instruction format (in control-flow graph) can be found in Figure~\ref{fig:openmp-llvm}(c).
As we can see, while this simple loop has only 5 basic blocks in serial implementation, even with a canonical loop format, its OpenMP version results in 12 basic blocks.
After characterizing each instruction in detail, we find three main reasons.
\textit{(1) Loop Transformation}: In order to execute programs in parallel, OpenMP transformed the loop into a canonical form suitable for runtime scheduling, where loop bounds, strides, and iteration variables are managed explicitly, resulting in more instructions and more basic blocks.
\textit{(2) Support for Edge Cases}: OpenMP must deal with edge cases like empty loops or loops with unusual bounds, adding conditional branches to check for these cases.
\textit{(3) Parallel Execution Support}: OpenMP handles shared and thread-private variables using different strategies, including parallel, serial, and atomic computations, often relying on external libraries to ensure correct execution.
Although (1) and (2) introduce additional instructions, they always execute in a thread-safe manner and have minimal impact on instruction duplication. Consequently, a key requirement for instruction duplication in OpenMP is to handle (3) properly.
\begin{boxH}
\textbf{Insight I}:
The key requirement for instruction duplication in OpenMP programs is managing parallel computation patterns, such as shared variables and atomic operations.
\end{boxH}

\vspace{-5mm}

In real-world scenarios, users may utilize diverse OpenMP APIs in complex ways, such as frequent synchronization, master/slave thread divergence, and nested directives.
Figure~\ref{fig:openmp-llvm}(b) illustrates the parallel region of the Rodinia~\cite{che2009rodinia} Lud benchmark. As seen, multiple directives are applied at different loop levels, alongside synchronization barriers, making it difficult to distinguish parallel and serial regions at the source code level.
However, after we compile this code into instructions, we have two key observations.
(1) LLVM compiler only treats the \textit{outermost OpenMP directive} as a separate function call, making it clear to identify parallel regions.
(2) Nested directives follow similar computation strategies (e.g. synchronization barrier and atomic operations) to its outermost one. 
\begin{boxH}
\textbf{Insight II}:
We distinguish parallel and serial regions in OpenMP programs by identifying the outermost directives. In this work, we define an outermost OpenMP directive as an \textbf{OpenMP Kernel} for simplicity.
\end{boxH}

\subsection{Dynamic Prediction for Programs via LLM}
\label{sec:initial-study-llm}
Recall that the other challenge is performance-aware error detection.
Thread count has a significant impact on the runtime behavior of parallel programs, even without any fault tolerance mechanism (see Figure~\ref{fig:challenge-2-1}). When combined with instruction duplication, this impact becomes even more unpredictable~\cite{huang2023characterizing}.
A promising error detection technique must therefore introduce replicas in a way that maintains both fault coverage and minimal runtime overhead, ideally by selecting the best thread configuration.
While dynamic profiling could help identify such configurations, it is often expensive in real-world workloads~\cite{cavelan2019detection,raut2020evaluating,huang2022mitigating}.
To avoid this cost, we seek a solution that relies solely on compile-time analysis.
Recently, large language models (LLMs) have shown strong potential for code analysis, such as compiler optimization~\cite{cummins2023large} and dynamic profiling~\cite{chatterjee2024phaedrus}, making them a promising tool in our scenario.


To evaluate the potential of LLMs for dynamic analysis in parallel programs, we conduct an initial study using \textit{Phi-3.5-mini-instruct}, a lightweight LLM with 3.82 B parameters released by Microsoft~\cite{abdin2024phi}.
Compared to its larger counterpart, Phi-3.5-MoE, it achieves competitive performance on code tasks (61.5 vs. 70.7) while requiring 15$\times$ fewer parameters, making it feasible for local inference on a single GPU.
In our setup, we extract the code corresponding to each program's parallel region as the initial prompt.
We then provide runtime data for 1, 2, and 4 threads, and ask the model to predict runtimes for 8 and 16 threads.
Initially, we found that predictions based on normalized runtime performed poorly, so we switched to raw runtime values (in seconds).
For clarity, we still report results in normalized format (with the serial runtime set to 1).
We reuse the 8 parallel programs from Rodinia, as in Section~\ref{sec:understanding}.

\begin{figure}[h]
\centering
\includegraphics[width=1.0\linewidth]{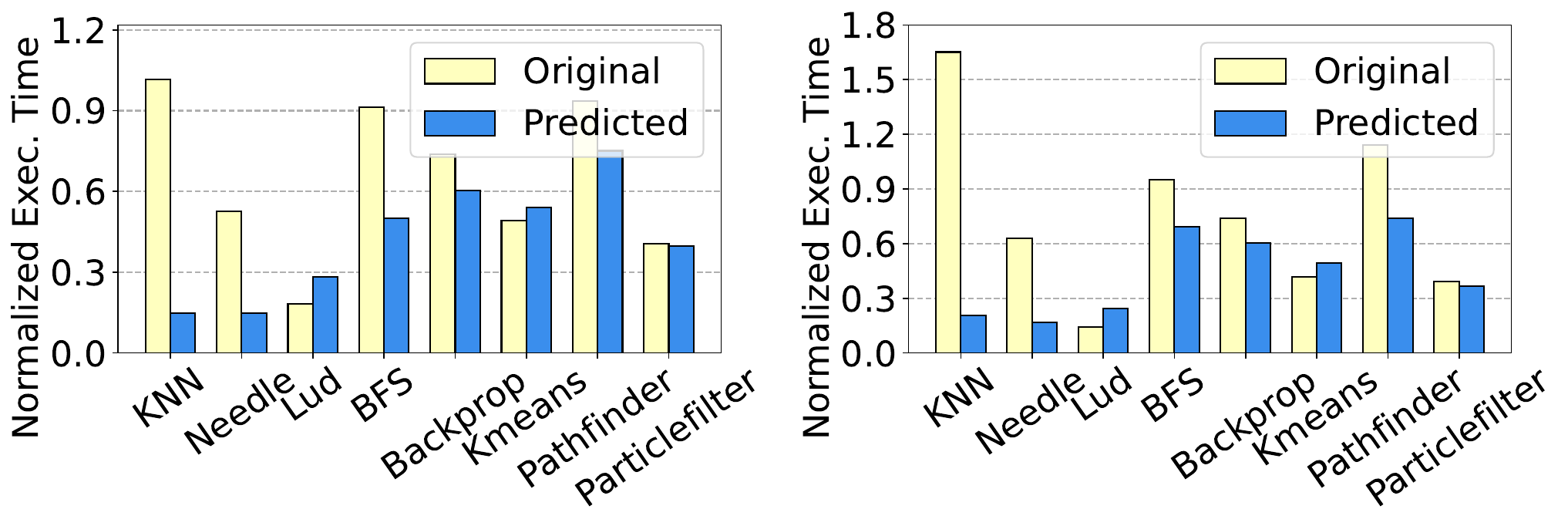}
\vspace{-8mm}
\caption{Original vs LLM predicted parallel program execution time for thread count 8 (left) and 16 (right).}
\label{fig:dynamic-llm}
\end{figure}

Figure~\ref{fig:dynamic-llm} presents the results.
At first glance, except Backprop and Particlefilter, the LLM struggles with accurate runtime prediction for most programs.
For instance, at 8 threads, KNN has an actual normalized runtime of 1.01 (i.e., no speedup), but the LLM incorrectly predicts 0.147 (suggesting an 8$\times$ speedup).
However, \textit{the LLM consistently gets it right while predicting trends} of performance change from 8 to 16 threads.
Interestingly, the LLM also provides rationales alongside its predictions.
For example, in BFS, it correctly anticipates a slowdown when increasing from 8 to 16 threads, citing the overhead of implicit barriers in loop-based parallelism.
The only exception is Pathfinder.
We find that Pathfinder's parallel region contains only a short 7-line \verb|for| loop, with most of the execution time dominated by serial code and output operations, which are difficult for LLMs to reason about.
However, once this context is explicitly included, LLM can produce accurate predictions: consistent performance across thread counts within Pathfinder.

\begin{boxH}
\textbf{Insight III}:
While LLMs may not predict exact runtimes, they can infer performance trends from code based on prior knowledge.
Furthermore, prompt refinement significantly improves prediction quality in challenging cases.
\end{boxH}

\section{Our Solution: \tech}

In this work, we propose \tech (\textit{\textbf{PaR}allel \textbf{I}nstruction \textbf{D}uplication}), a soft error detection framework for OpenMP-based parallel software.
\tech operates entirely at compile-time, requiring only static code transformation and a few lightweight, device-side LLM inferences, while maintaining high error detection effectiveness and low runtime overhead.
Given an arbitrary parallel program, \tech performs two independent steps to generate a protected executable binary.
First, after compiling the source code to LLVM IR, \tech applies Parallel-aware Code Transformation (\ding{202}) to produce replica-instrumented IR that is compatible with both serial and parallel regions, addressing \textbf{C-1}.
Second, based on an offline characterization study, \tech utilizes learned prompts to guide the LLM-tuned Performance Modeling (\ding{203}) in inferring optimal thread settings for each OpenMP kernel, avoiding expensive dynamic profiling and addressing \textbf{C-2}.
The entire \tech pipeline is fully automated through Python scripts, requiring no user intervention, and can be seamlessly integrated into production workflows.
In the following context, we will explain both \ding{202} and \ding{203} in detail.

\begin{figure}[h]
\centering
\includegraphics[width=1.0\linewidth]{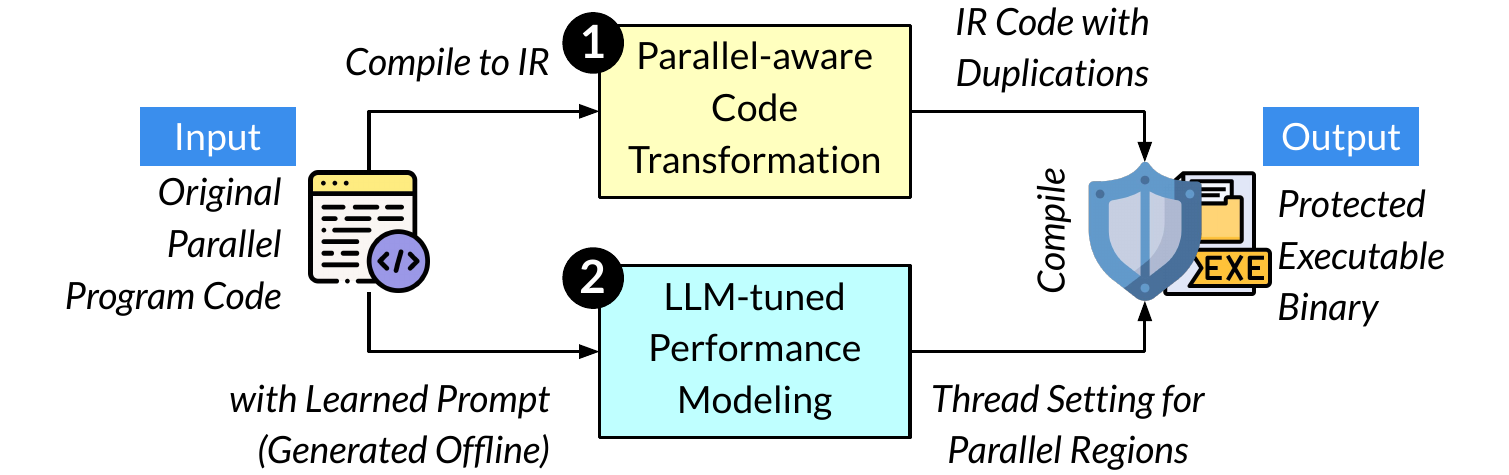}
\vspace{-3mm}
\caption{Workflow of \tech framework.}
\label{fig:ompid-framework}
\end{figure}



\subsection{Parallel-aware Code Transformation (\ding{202})}
\label{sec:high-level-ompid}

\begin{figure}[h]
\centering
\includegraphics[width=1.0\linewidth]{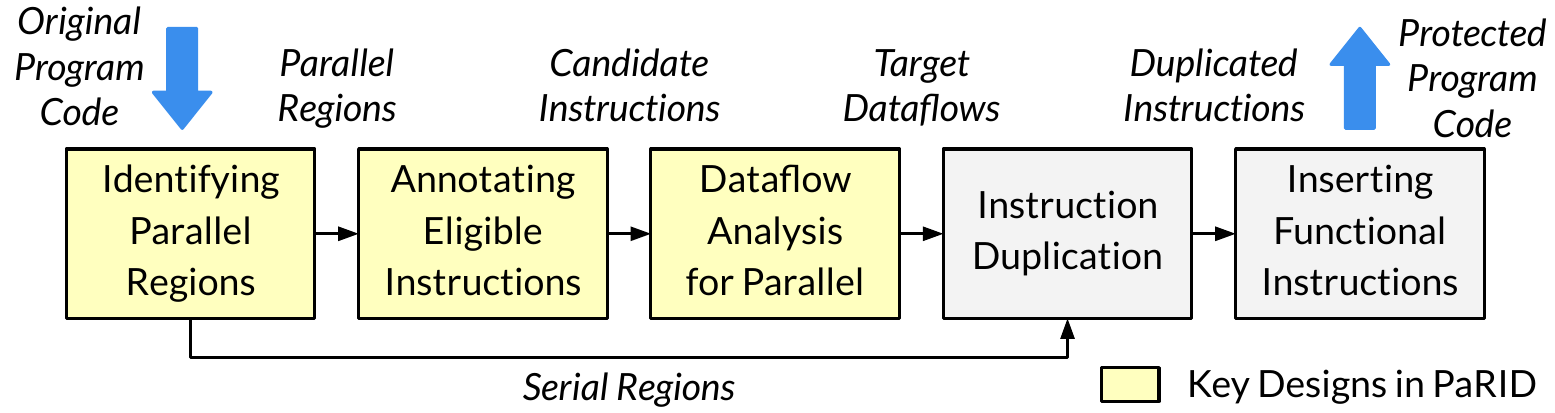}
\vspace{-3mm}
\caption{Workflow of Parallel-aware Code Transformation in \tech, where the input and output are both in LLVM IR.}
\label{fig:ompid-workflow}
\end{figure}

Figure~\ref{fig:ompid-workflow} illustrates how Parallel-aware Code Transformation in \tech instruments duplication along with functional instructions in a thread-safe manner.
Given the original program code in IR, \tech first performs \textit{Identifying of Parallel Regions} by localizing each OpenMP kernel (as described in \textbf{Insight II}, Section~\ref{sec:understanding}). At the instruction level, this is achieved by matching the keyword {\small\texttt{omp\_outlined}} in function names.
For serial regions, \tech downgrades to traditional instruction duplication, directly instrumenting both duplicated and functional instructions\footnote{Functional instructions in instruction duplication (see Figure~\ref{fig:instruction-duplication}) refer to the ones inserted to handle mismatch detection (e.g., comparisons and conditional branches), error reporting (e.g., function calls), and optional recovery triggering~\cite{lu2014sdctune,laguna2016ipas,huang2023characterizing}.}. 
For parallel regions, \tech conducts \textit{Annotation of Eligible Instructions} and \textit{Dataflow Analysis for Parallel Regions}. 
These two steps enable \tech to properly handle parallel computation patterns (as mentioned in \textbf{Insight I}, Section~\ref{sec:understanding}).
Finally, with target dataflows contained by candidate instructions, \tech instruments the replicas and functional instructions, generating the protected IR.
We implement the entire code transformation within a single LLVM pass, ensuring the usability.

After identifying parallel regions, the design of \tech for handling OpenMP parallel regions focuses on two key components: \textit{Annotating Eligible Instructions} and \textit{Dataflow Analysis for Parallel Regions}. The former prunes ineligible instructions and appropriately annotates and processes OpenMP-specific instructions, whereas the latter extracts program dataflows in a thread-safe manner. Together, these components facilitate the subsequent steps of instruction duplication and insertion of functional instructions for OpenMP.


\subsubsection{Annotating Eligible Instructions}
In addition to computational instructions (identical to those in the serial implementation), LLVM generates extra parallel-supporting instructions in each OpenMP kernel.
As noted in Section~\ref{sec:understanding}, instructions related to loop transformations (e.g., {\small\texttt{\%.omp.stride}}) and edge cases supporting (e.g., {\small\texttt{\%omp.precond}}) are thread-safe. 
Among the remaining instructions, we identify two types that are not eligible for duplication.
\textit{(1) Write to Shared Region}. Parallel access to a shared array is typically translated into {\small\texttt{store}} operations targeting independent indices.
In \tech, these {\small\texttt{store}} operations are excluded from duplication. We reckon this exclusion is justified, as the values being stored are already duplicated and verified in thread-preserved regions prior to storing. 
Indiscriminately duplicating these write-to-memory instructions could disrupt memory consistency or lead to unnecessary overhead, especially when handling shared memory. Furthermore, our fault model assumes that memory errors are corrected by hardware mechanisms such as ECC, rendering additional duplication redundant for these operations.
\textit{(2) OpenMP Runtime Calls}.
We also prune such function calls, such as {\small\texttt{kmpc\_barrier}} for synchronization and {\small\texttt{kmpc\_critical}} for managing critical regions, for duplication.
These runtime calls do not directly impact computation or dataflow, which requires verifying error occurrences.
Instead, they act as control mechanisms to ensure proper execution order and mutual exclusion among threads~\cite{dagum1998openmp}.
Duplicating such calls could interfere with the correct functioning of the OpenMP runtime.


\begin{figure}[h]
\centering
\includegraphics[width=1.0\linewidth]{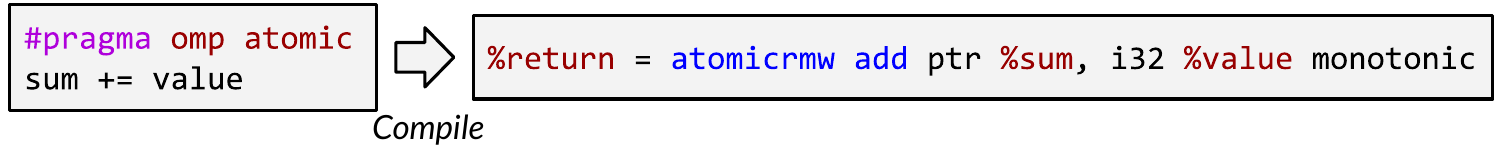}
\vspace{-5mm}
\caption{Atomic operations at LLVM IR.}
\label{fig:atomicrmw}
\end{figure}

Note that some atomic operations in OpenMP programs reflect as an {\small\texttt{atomicrmw}} instruction.
One example is shown in Figure~\ref{fig:atomicrmw}.
In this case, multiple threads perform \textit{read-modify-write} operations ({\small\texttt{add \%value}}) to the same shared address ({\small\texttt{\%sum}}), where {\small\texttt{return}} register denotes the value of shared register before the computation.
We do not duplicate such instructions because doing so would result in mismatched execution orders between two copies and a false error report. 
Instead, we apply a pre-checking mechanism for all operands involved in the {\small\texttt{atomicrmw}} instruction, ensuring values passed in is error-free. Specifically, an additional comparison instruction is inserted to validate the operands before the atomic operation executes.
While this approach slightly breaks the dataflow and incurs minor additional overhead, it is justified because {\small\texttt{atomicrmw}} operations are relatively infrequent in most workloads~\cite{che2009rodinia}. 
Most importantly, it ensures no loss of fault coverage effectiveness in \tech.


\subsubsection{Dataflow Analysis for Parallel Region}
\label{sec:dataflow-analysis}
In \tech, we perform duplication and instrument a functional comparison instruction ({\small\texttt{icmp}}) at the end of each \underline{s}tatic \underline{d}ata \underline{d}ependency \underline{s}equence (SDDS), defined as a chain of instructions where the output of one instruction is used as an operand by another~\cite{petersen1996static,li2018modeling}.
This approach avoids performing comparisons after every pair of duplicated instructions, which would introduce extra jumps and disrupt the program's control flow.
In OpenMP programs, implementing this requires \tech to analyze program dataflow and ensure that each SDDS is identified in a thread-safe manner. 
The key strategy in this step is to analyze instruction operand usage while also extracting each SDDS immediately before accessing a shared region, particularly within \textit{critical} sections, where only one thread executes simultaneously.

\begin{figure}[h]
\centering
\includegraphics[width=1.0\linewidth]{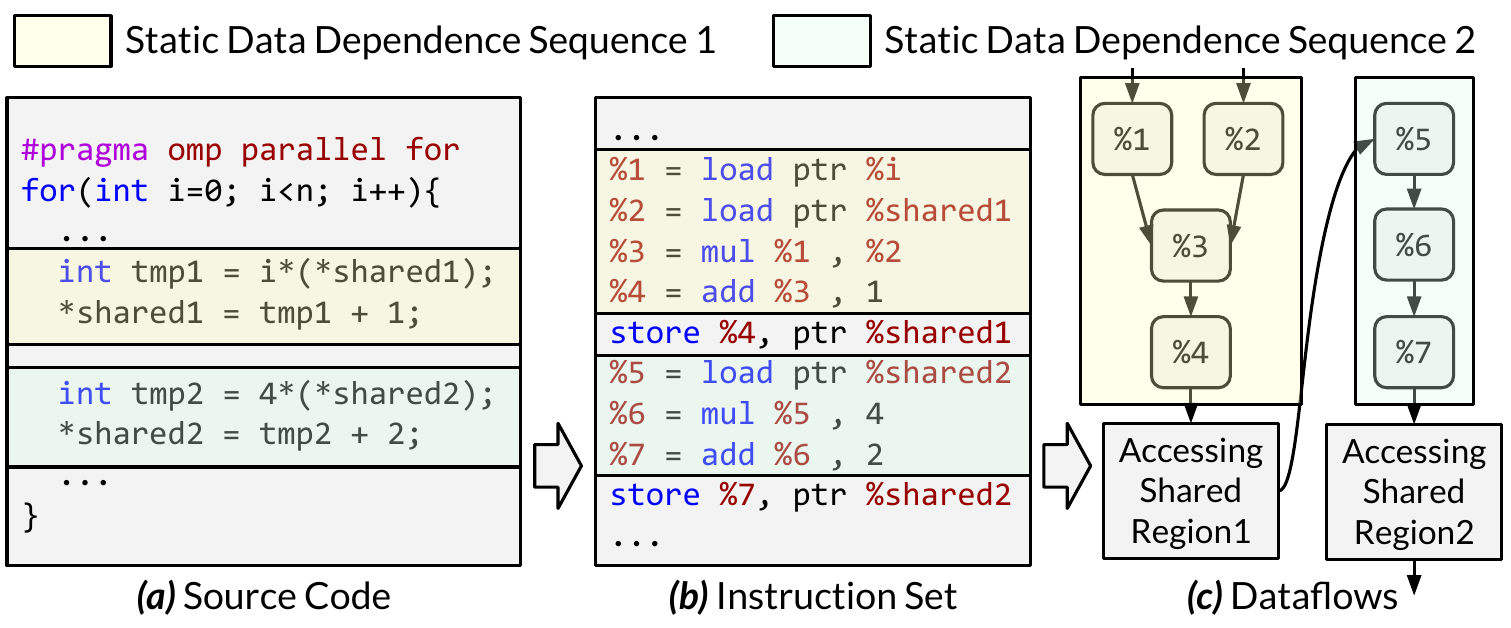}
\vspace{-3mm}
\caption{Illustrating Dataflow Analysis for Parallel Regions.}
\label{fig:static-dds}
\end{figure}

Figure~\ref{fig:static-dds} illustrates how \tech recognizes SDDSs in an OpenMP parallel {\small\texttt{for}} loop.
After compiling the source code (Figure~\ref{fig:static-dds}(a)) into its instruction set representation (Figure~\ref{fig:static-dds}(b)), the basic block is divided into two distinct SDDSs, each concluding with a {\small\texttt{store}} operation to write data to a shared region. At the dataflow level (Figure~\ref{fig:static-dds}(c)), \tech will duplicate each SDDS and compare the results between {\small\texttt{\%4}} and {\small\texttt{\%4'}} as well as {\small\texttt{\%7}} and {\small\texttt{\%7'}} in later instrumentation steps, ensuring thread safety while effectively detecting errors.
Since LLVM IR is in SSA form~\cite{lattner2004llvm}, operand tracing and SDDS extraction can be done efficiently using def-use chains~\cite{kennedy1978use}.



\subsection{LLM-tuned Performance Modeling (\ding{203})}

\begin{figure}[h]
\centering
\includegraphics[width=1.0\linewidth]{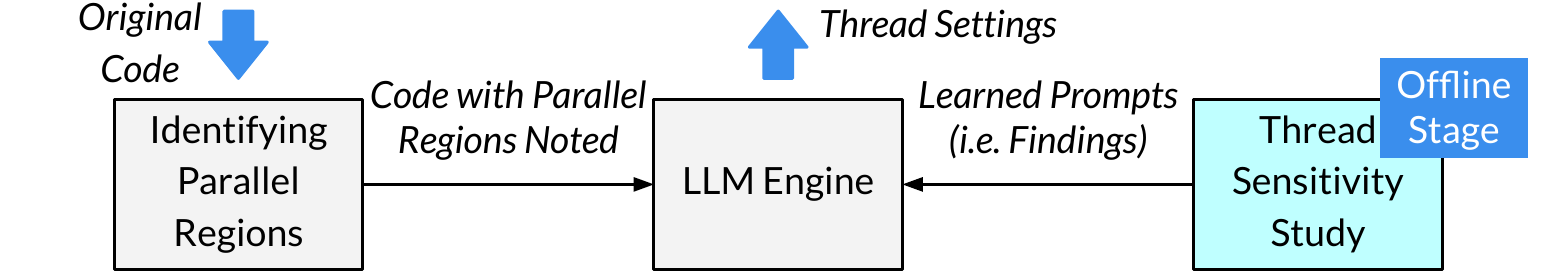}
\vspace{-5mm}
\caption{Workflow of LLM-tuned Performance Modeling.}
\label{fig:llm-workflow}
\end{figure}

\begin{figure*}[ht!]
    \centering
    \subfigure[Hotspot3D]{
        \includegraphics[width=0.13\textwidth]{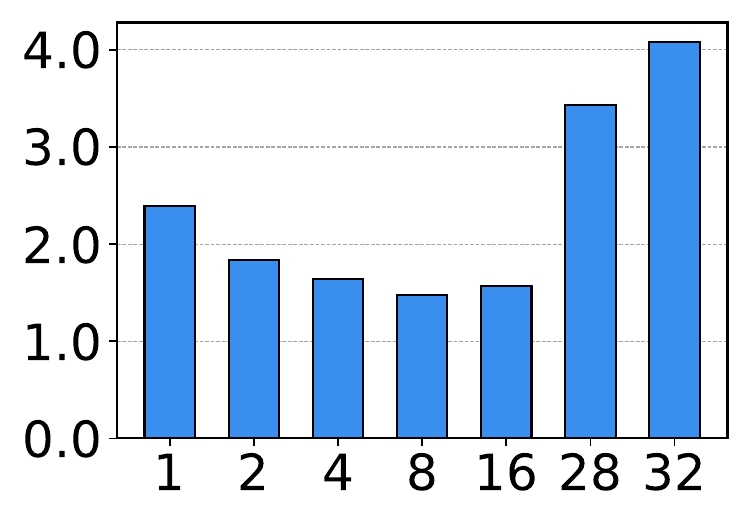}
    }
    \subfigure[Srad]{
        \includegraphics[width=0.13\textwidth]{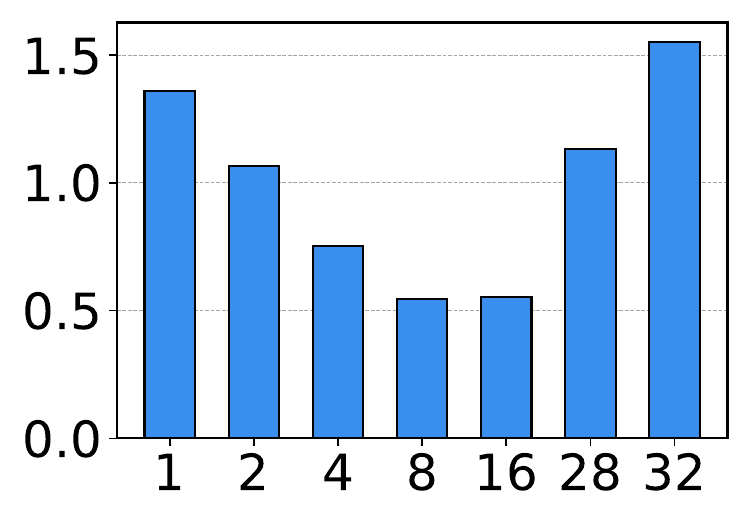}
    }
    \subfigure[LavaMD]{
        \includegraphics[width=0.13\textwidth]{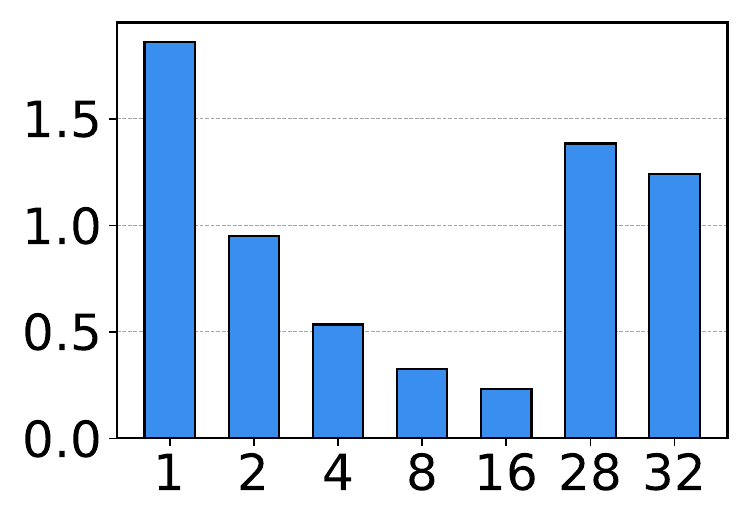}
    }
    \subfigure[KNN]{
        \includegraphics[width=0.13\textwidth]{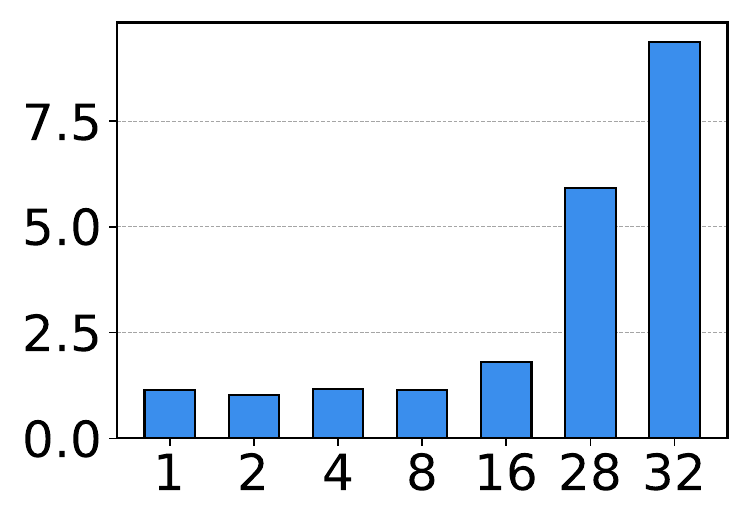}
    }
    \subfigure[Needle]{
        \includegraphics[width=0.13\textwidth]{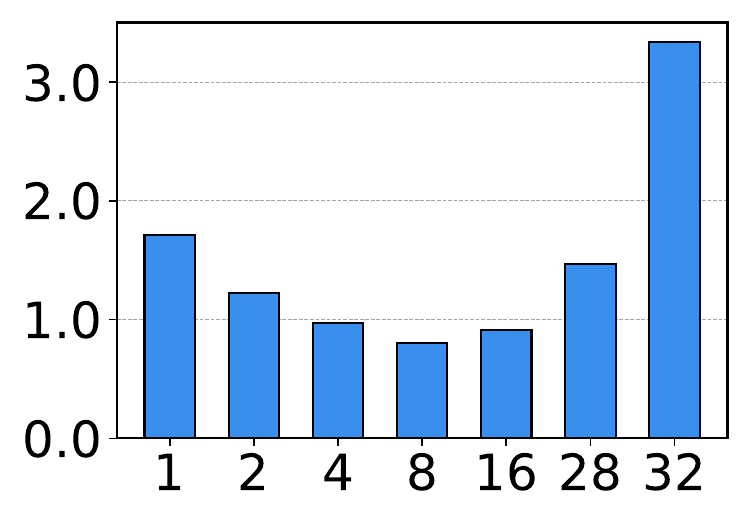}
    }
    \subfigure[Hotspot]{
        \includegraphics[width=0.13\textwidth]{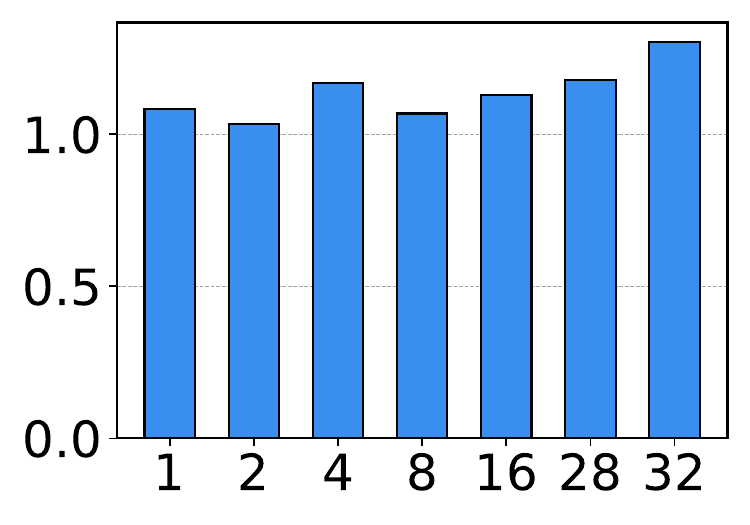}
    }
    \subfigure[Lud]{
        \includegraphics[width=0.13\textwidth]{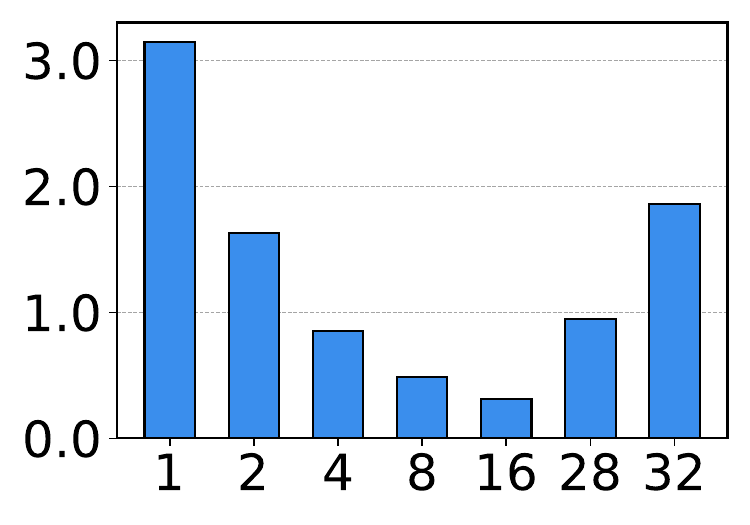}
    }\\

    \subfigure[BFS]{
        \includegraphics[width=0.13\textwidth]{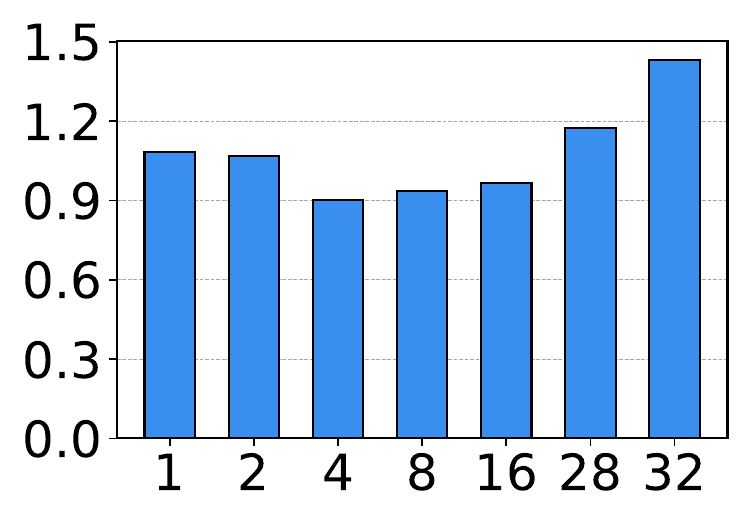}
    }
    \subfigure[B+Tree]{
        \includegraphics[width=0.13\textwidth]{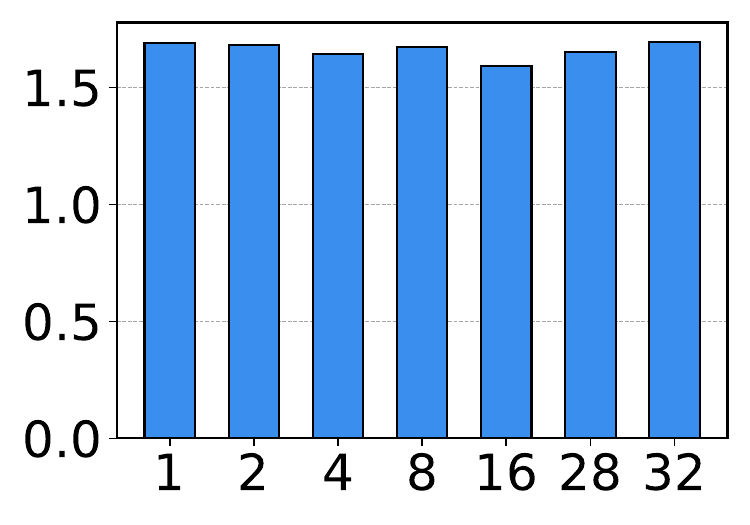}
    }
    \subfigure[Backprop]{
        \includegraphics[width=0.13\textwidth]{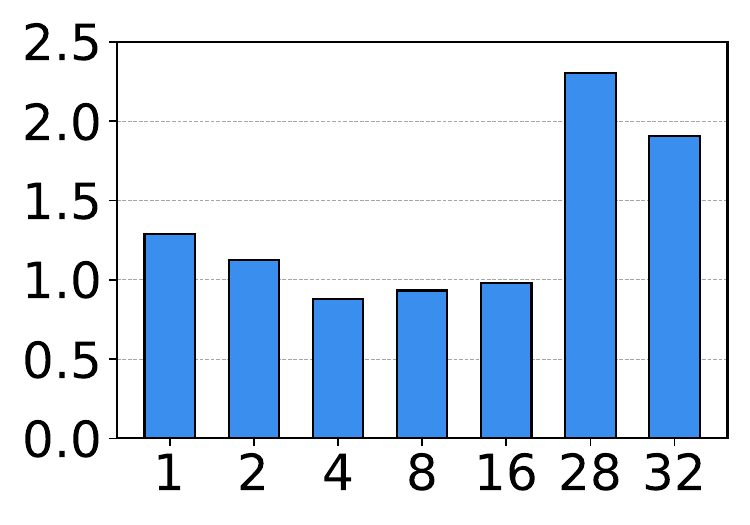}
    }
    \subfigure[Myocyte]{
        \includegraphics[width=0.13\textwidth]{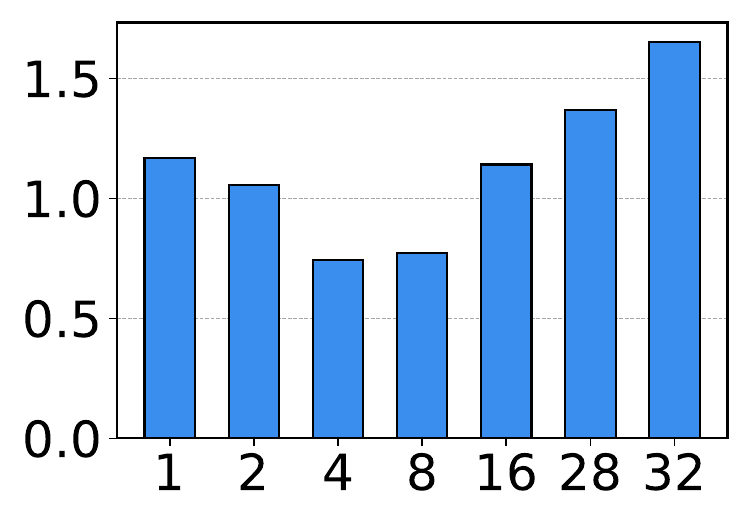}
    }
    \subfigure[Kmeans]{
        \includegraphics[width=0.13\textwidth]{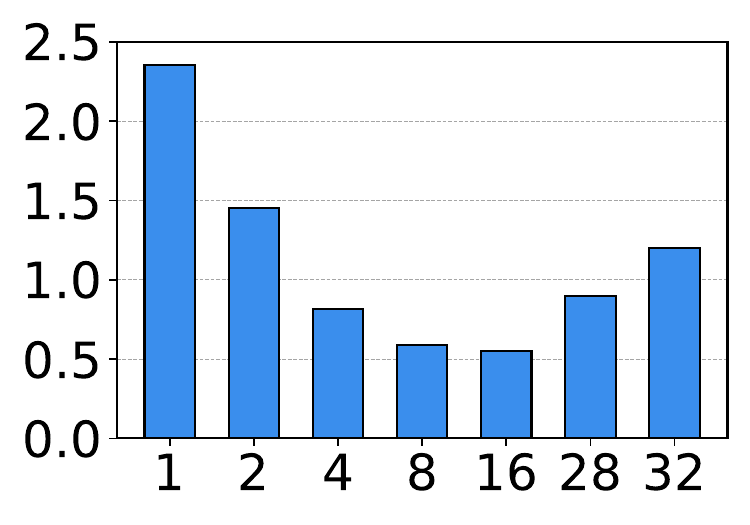}
    }
    \subfigure[Heartwall]{
        \includegraphics[width=0.13\textwidth]{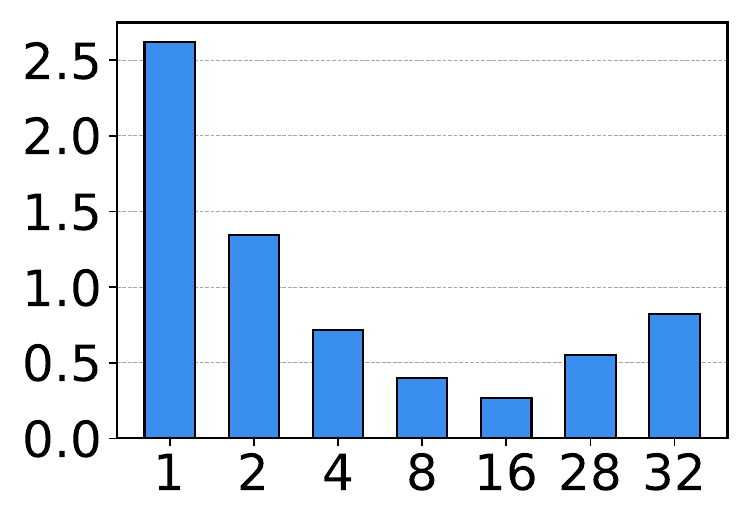}
    }
    \subfigure[Particlefilter]{
        \includegraphics[width=0.13\textwidth]{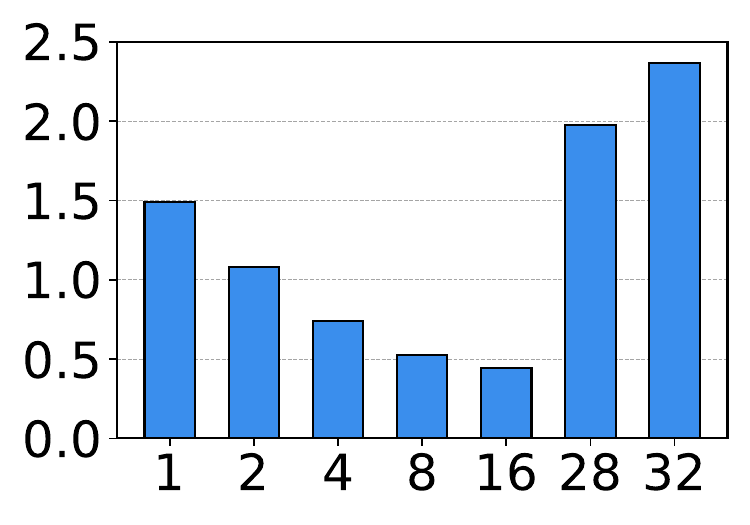}
    }

    \vspace{-2mm}
    \caption{Thread Sensitivity Study using Rodinia benchmarks with \tech~\ding{202} transformation. For each benchmark, x-axis is the number of threads; y-axis is the normalized runtime. Pathfinder is excluded since it is dominated by CPU computations.}
    \label{fig:thread-sensitivity-study}
\end{figure*}

Recall that parallelism is inherently introduced to accelerate program execution~\cite{gustafson1988reevaluating,skinner2005performance}.
As such, \tech must reduce the runtime overhead introduced by error detection mechanisms (i.e., \ding{202}) to preserve the performance benefits of parallelism.
Both thread count (see Figure~\ref{fig:challenge-2-1}) and instruction duplication overhead (see Figure~\ref{fig:challenge-2-2}) impact runtime differently across programs and regions.
A natural solution is to identify the optimal thread setting for each parallel region, reducing performance loss.
Additionally, to ensure usability in real-world settings, this tuning should be done entirely at compile time.
Inspired by \textbf{Insight III} in Section~\ref{sec:initial-study-llm}, we leverage the potential of LLMs to infer dynamic performance characteristics from code, especially when guided by well-constructed prompts.
\tech adopts this strategy through LLM-Tuned Performance Modeling, which automatically selects the best thread configuration for each OpenMP kernel in the input program.

The idea of this step is as follows: \textit{We transform a set of widely used parallel benchmarks using \ding{202} in \tech, conduct a detailed thread-level characterization study, and extract key findings to serve as prompts for guiding LLM inference}.
Figure~\ref{fig:llm-workflow} illustrates the workflow of LLM-Tuned Performance Modeling in \tech.
In the offline stage, we perform a one-time \textit{Thread Sensitivity Study} across 15 Rodinia benchmarks~\cite{che2009rodinia} to identify performance trends under different thread counts.
From this study, we extract several generalizable findings, which are encoded as prompt components for the LLM.
In the online stage, given a new input program, \tech first applies \textit{Identifying Parallel Regions} (as in \ding{202}) to locate each OpenMP kernel.
Then, the \textit{LLM Engine} analyzes each kernel and recommends a thread configuration based on the previously learned prompt strategies.
By doing so, \tech shifts the expensive dynamic analysis to the offline phase, making online prediction both lightweight and accurate.
We describe the details of the \textit{Thread Sensitivity Study}.

\subsubsection{Setups before Characterization}
We use 15 benchmarks from the Rodinia suite~\cite{che2009rodinia} (v3.1)~\cite{rodinia-3}, which provides both serial and parallel implementations and covers diverse computation patterns across multiple domains.
There are 19 benchmarks in total, and we try to include all of them. However, we exclude 4 benchmarks due to incompatibility with our toolset. For example, current LLVM-OpenMP project~\cite{llvm-openmp} does not support the required cross-machine dependencies in Leukocyte.
Rodinia is particularly well-suited for our study, as most of its benchmarks are lightweight and contain either a single OpenMP kernel or multiple structurally similar kernels, simplifying analysis.
To evaluate the impact of instruction duplication, we report \textit{normalized runtime}, where the baseline is the serial execution of the unprotected program (set to 1.0).
Unlike prior works~\cite{kalra2020armorall,huang2023characterizing,laguna2016ipas}, we do not use the traditional overhead metric (i.e., Protection/Original under the same thread count), as changes in thread configuration can naturally reduce runtime.
In such cases, a naive overhead comparison may misleadingly suggest improved performance even when duplication introduces cost.
All experiments are conducted on a 28-core machine.

\subsubsection{Thread Sensitivity Study}
\label{sec:thread-sensitivity-study}
Figure~\ref{fig:thread-sensitivity-study} presents the results of \tech Thread Sensitivity Study, demonstrating \tech~\ding{202} can be seamlessly integrated for parallel programs.
We report results for 14/15 benchmarks; Pathfinder is excluded, as its runtime is dominated by serial computations (as shown in Section~\ref{sec:initial-study-llm}).
All benchmarks use the standard input provided by Rodinia~\cite{rodinia-3}.
For each benchmark, we report normalized execution time under thread counts \{1, 2, 4, 8, 16, 28, 32\}, where 28 corresponds to the machine core count.
We include 32 threads to observe the impact of context switching (i.e., oversubscription~\cite{iancu2010oversubscription}).
Rather than identifying the "best" thread count for each benchmark, which is highly machine-specific, we extract generalizable findings from the data, guiding LLM inference in a portable and system-agnostic manner.

\textit{\textbf{Finding 1}: If a benchmark has minimal OpenMP regions and large serial sections, then it will exhibit consistent runtime across thread settings.} 
Besides Pathfinder, it is also observed in B+Tree, where only two \verb|for| loops are parallelized, while $\sim$2,500 lines of complex loop structures remain in the serial region. As a result, \tech exhibits nearly identical execution time across thread counts.
While long serial code may challenge customer-side LLM inference, this property can be easily identified through simple static analysis using LLVM passes.
In contrast, Hotspot also shows consistent runtime across threads, but the cause is different: its standard input is too small, making the runtime dominated by serial preprocessing and postprocessing.
However, such input size issues are less relevant in real-world workloads; we explore this further in Section~\ref{sec:new-evaluation}.

\begin{figure}[h]
\centering
\includegraphics[width=1.0\linewidth]{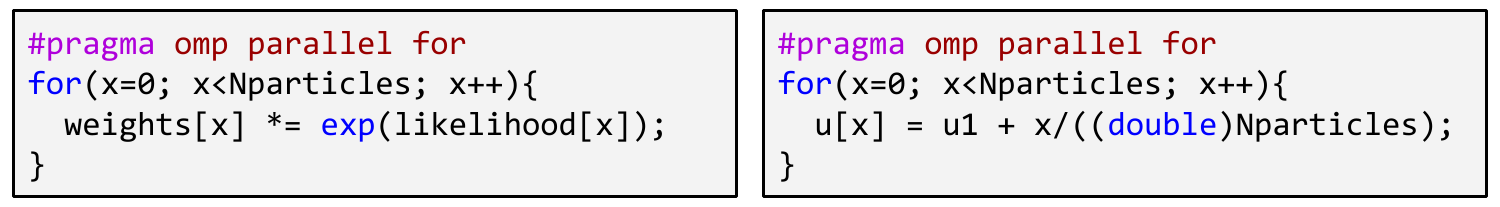}
\vspace{-5mm}
\caption{Illustrating two embarrassingly parallel OpenMP kernels (by \texttt{\small{\#pragma omp parallel for}}) in Particlefilter.}
\label{fig:particlefilter}
\end{figure}

\textit{\textbf{Finding 2}: If the OpenMP kernel is embarrassingly parallel (e.g., for loop with no dependencies), then higher thread counts can be applied for \tech.}
This is a commonly observed pattern, seen in Heartwall, LavaMD, Lud, Kmeans, and Particlefilter, all of which achieve peak performance with 16 threads.
These benchmarks follow a data-parallel model~\cite{hillis1986data}, distributing loop iterations evenly across threads without requiring explicit thread boundaries.
Figure~\ref{fig:particlefilter} shows two examples of embarrassingly parallel OpenMP kernels.
In Lud, although the code features more complex or interleaved control flow, \tech's \textit{Identifying Parallel Regions} step can still easily recognize these as embarrassingly parallel kernels.

\textit{\textbf{Finding 3}: If thread count nears the core count,
then \tech’s per-thread workload causes the runtime to increase (especially from complicated control-flow).} 
This behavior is consistently observed in 13/15 benchmarks.
Recall in Section~\ref{sec:understanding}, we find that OpenMP naturally introduces additional control-flow and parallel-support instructions.
To detect soft errors, \tech~\ding{202} further inserts replicated and functional control-flow operations, resulting in performance degradation, especially approaching the hardware core limit.

\textit{\textbf{Finding 4}: If the OpenMP kernel has frequent synchronizations, then \tech runs best with moderate threads.} 
This behavior is observed in KNN.
When an OpenMP kernel includes frequent barriers and critical regions, thread-level waiting becomes unavoidable.
With increased per-thread workload from duplication, a moderate thread count achieves better hardware utilization.

\textit{\textbf{Finding 5}: If the OpenMP kernel uses many variables, then \tech causes high register and memory pressure. Moderate threads avoid slowdown.}
This is observed in Myocyte and Needle.
For example, Myocyte's function, \verb|ecc()|, uses over 600 scalar variables.
Since \tech nearly doubles the number of registers required per thread for performing error detection, this leads to register spills and memory stalls.
Therefore, even if these kernels may have balanced computations, a moderate thread number yields better performance.
 
\textit{\textbf{Finding 6}: If the algorithm needs power-of-2 partitioning, then use power-of-2 thread counts. Irregular counts in \tech cause more control-flow and overhead.} This is observed in Backprop, where performance degrades significantly with 28 threads but remains acceptable with 16 and 32 threads.
The algorithm computes partial derivatives across layers, each sized as a power of two, and relies on even partitioning.
Using 28 threads introduces boundary handling issues, which increase control-flow complexity under \tech and result in higher overhead.
For such algorithms, thread counts should align with power-of-2 partitioning to avoid performance penalties.

\textit{\textbf{Finding 7}: \tech is sensitive to oversubscription. When the OpenMP kernel has load imbalance, runtime spikes when threads exceed core count.} 
While oversubscribing threads is a common technique to improve performance in OpenMP programs~\cite{yan2016proposal}, we find that \tech-transformed programs are consistently sensitive to this practice across all benchmarks.
In particular, when the OpenMP kernel contains imbalanced control flow, such as an outermost \verb|if-else| with imbalanced workloads, performance degrades significantly.

\textit{\textbf{Finding 8}: If a program has multiple OpenMP kernels, then \tech performs worse near core-limit.} 
This is observed in Particlefilter, where \tech's performance degrades starting around 18 threads (not shown in Figure~\ref{fig:thread-sensitivity-study}).
Although all 10 OpenMP kernels in this benchmark are embarrassingly parallel, they are interleaved with serial regions, introducing implicit synchronization overhead.
Under high thread counts, this overhead is amplified, and \tech's per-thread workload exacerbates the issue.
In such cases, using fewer than the core-limit threads yields better performance.

\subsubsection{Prompt Construction}
Beyond natural language, the LLM Engine requires four core components to form the complete prompt:
\textit{(1) Machine Specification:}
The LLM Engine needs access to CPU specifications, particularly the core limit, to help produce generalizable recommendations.
\textit{(2) Parallel Code Region:}
Each OpenMP kernel is provided as input for the LLM to infer its best-suited thread configuration.
\textit{(3) Learned Prompts (Findings):}
These are offline-refined insights derived from our thread sensitivity study.
Most are encoded as \textbf{if-then} rules~\cite{qiao2021learning,wang2024can} (except \textit{Finding 7}) to improve LLM interpretability.
\textit{(4) Serial Code Regions:}
The LLM is also given a broader context of the program, including serial regions, which is essential for applying \textit{Finding 1} and \textit{Finding 8}.
However, in real-world scenarios, the entire codebase may be prohibitively large (e.g., thousands of lines in NPB~\cite{bailey2010parallel}), exceeding the capacity of lightweight LLMs.
To address this, we offload parts of the analysis to static LLVM passes (such as counting OpenMP kernels for \textit{Finding 8}) to reduce inference workload while preserving key semantic information.
\section{Evaluation}
\label{sec:new-evaluation}

We provide evaluation setups and evaluate \tech in this section.

\subsection{Evaluation Setups}

\subsubsection{Benchmarks}
We select all eight original benchmarks in NAS Parallel Benchmarks (NPB) suite~\cite{bailey1991parallel} (V3.0~\cite{bailey2010parallel}).
Details are shown in Table~\ref{tab:benchmark}.
These programs are specifically designed to evaluate parallel performance and include OpenMP support.
Compared to other suites~\cite{guthaus2001mibench,stratton2012parboil,che2009rodinia}, NPB programs have larger workloads, tunable input classes, and diverse computation/memory patterns, such as irregular memory access in CG and embarrassingly parallel in EP, making them more reflective of real-world workloads.

\begin{table}[h]
\renewcommand{\arraystretch}{0.9}
\centering
\footnotesize
\begin{tabular}{c|l|c|r} \toprule
{\bf Name} & {\bf Description} & {\bf \# Kernels}  & {\bf SLOC}   \\ \midrule
IS & Integer sorting and ranking.               &  2   & 1,117\\
EP & Embarrassingly parallel Gaussian stats.    &  2   & 681\\
CG & Conjugate gradient linear system solver.   &  14  & 1,334\\
MG & Multi-Grid on a sequence of meshes.        &  10  & 1,690\\
FT & Discrete 3D fast Fourier Transform.        &  7   & 1,682\\
BT & Block Tri-diagonal solver.                 &  9   & 4,126\\
SP & Scalar Penta-diagonal solver.              &  7   & 3,482\\
LU & Lower-Upper Gauss-Seidel solver.           &  8   & 4,039\\
\bottomrule
\end{tabular}
\vspace{3mm}
\caption{Benchmark details. "\# Kernel" indicates the number of OpenMP kernels; "SLOC" indicates Source Lines of Code.}
\label{tab:benchmark}
\end{table}

\vspace{-3mm}
\subsubsection{Platform}
\label{sec:eva-plantform}
We conduct experiments on an Intel Xeon CPU (2.1 GHz) with 28 cores and 32 GB of RAM, running Ubuntu 20.04 OS. The pass for \tech is implemented using LLVM v15.0, which includes the standard optimizer and the Clang compiler. For parallel programs, we use OpenMP v5.0 (i.e. with {\small\texttt{\_OPENMP}} value 201811).
We use \textit{Phi-3.5-mini-instruct} LLM (3.82 B) from Microsoft~\cite{abdin2024phi} as the LLM Engine for \tech.
LLM Inference is conducted using an NVIDIA A100 GPU (40 GB, 108 SMs) with CUDA 12.6.

\subsubsection{Evaluation Methodology}
We evaluate \tech from two perspectives: execution time and fault coverage.
In this work, we prioritize execution time, as OpenMP parallelism is explicitly introduced to accelerate program execution~\cite{skinner2005performance}.
A key design goal of \tech is to use \ding{203} to mitigate the overhead introduced by \ding{202} through optimal thread tuning.
Consistent with Section~\ref{fig:thread-sensitivity-study}, we report \textit{normalized runtime}, relative to the serial version of the original, unprotected program (normalized to 1.0).
Evaluation settings for fault coverage are discussed separately in Section~\ref{sec:fault-coverage}.

For comparison, our baseline is the default thread configuration used in the NPB suite~\cite{bailey2010parallel}.
We also test alternative thread settings and confirm that this default yields reasonable performance.
To the best of our knowledge, there are no existing instruction duplication techniques specifically designed for parallel programs.
Additionally, we implement a serial instruction duplication baseline, representing state-of-the-art fault tolerance methods in non-parallelized programs~\cite{reis2005swift,mahmoud2018optimizing,huang2022mitigating,huang2024versatile,liao2025deep}.
This allows us to highlight the effectiveness of our parallel-aware duplication strategy in \tech~\ding{202}.



\subsection{Execution Time with \tech}

We evaluate \tech's execution time from following perspectives.

\begin{figure}[h]
\centering
\includegraphics[width=1.0\linewidth]{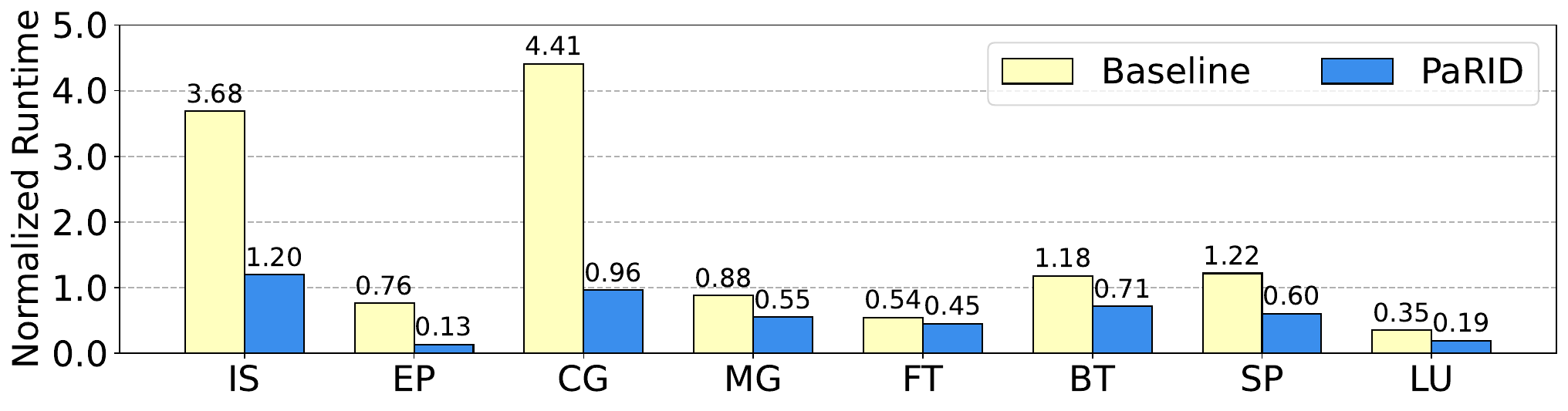}
\vspace{-6mm}
\caption{Overall execution time: \tech vs baseline.}
\label{fig:overall-result}
\end{figure}

\textbf{\textit{Overall Execution Time}}.
We first report the overall execution time comparison between \tech and the baseline method.
In the baseline setup, the NPB suite automatically uses the maximum available core count for execution, whereas \tech applies LLM-inferred thread counts for each OpenMP kernel.
We use the Class-A input (a standard test size) from the NPB suite~\cite{bailey2010parallel}.
Figure~\ref{fig:overall-result} presents the normalized runtime results.
In \tech, runtime varies from 0.12 in EP to 1.19 in IS, while in the baseline method, it ranges from 0.30 in LU to 4.40 in CG.
By selecting thread settings at compile time with LLM, \tech achieves an average 1.74$\times$ speedup over the baseline, with a maximum speedup of 4.94$\times$ on EP.

\begin{figure}[h]
    \centering
    \subfigure[IS]{
        \includegraphics[width=0.24\columnwidth]{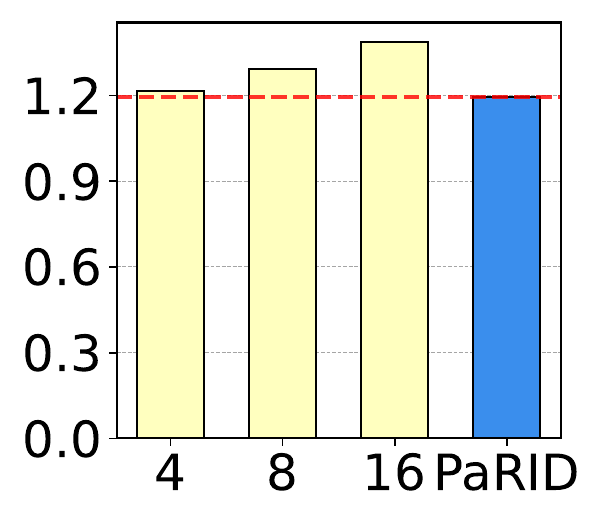}
    }
    \hspace{-3mm}
    \subfigure[EP]{
        \includegraphics[width=0.24\columnwidth]{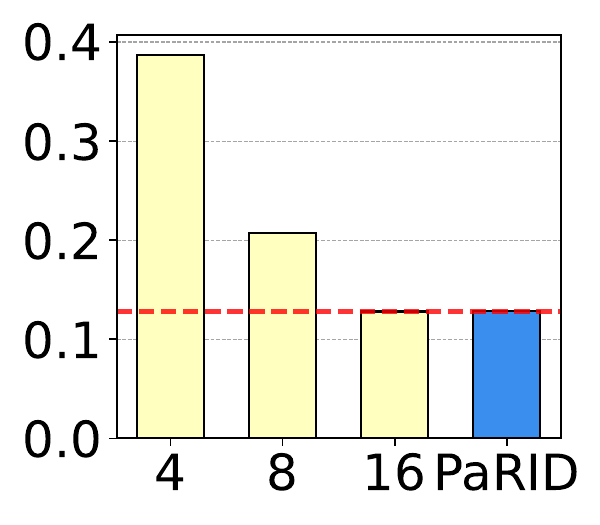}
    }
    \hspace{-3mm}
    \subfigure[CG]{
        \includegraphics[width=0.24\columnwidth]{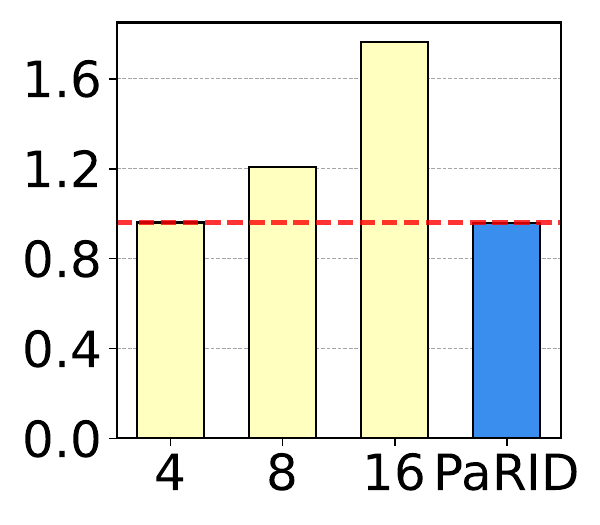}
    }
    \hspace{-3mm}
    \subfigure[MG]{
        \includegraphics[width=0.24\columnwidth]{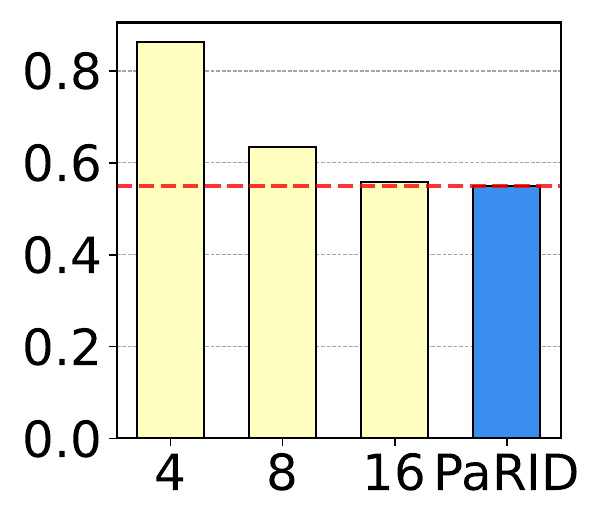}
    }

    \subfigure[FT]{
        \includegraphics[width=0.24\columnwidth]{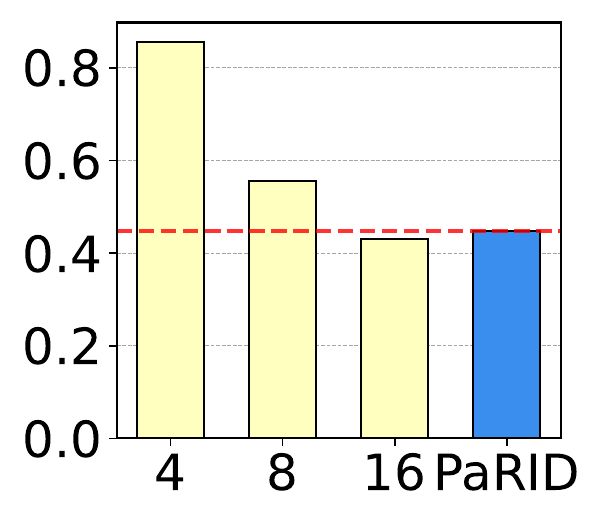}
    }
    \hspace{-3mm}
    \subfigure[BT]{
        \includegraphics[width=0.24\columnwidth]{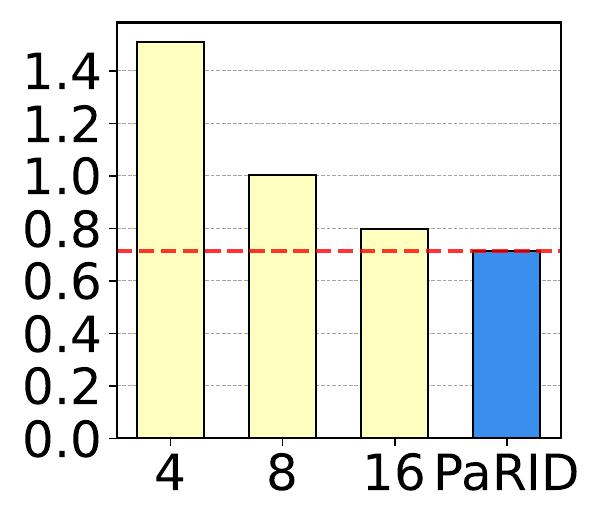}
    }
    \hspace{-3mm}
    \subfigure[SP]{
        \includegraphics[width=0.24\columnwidth]{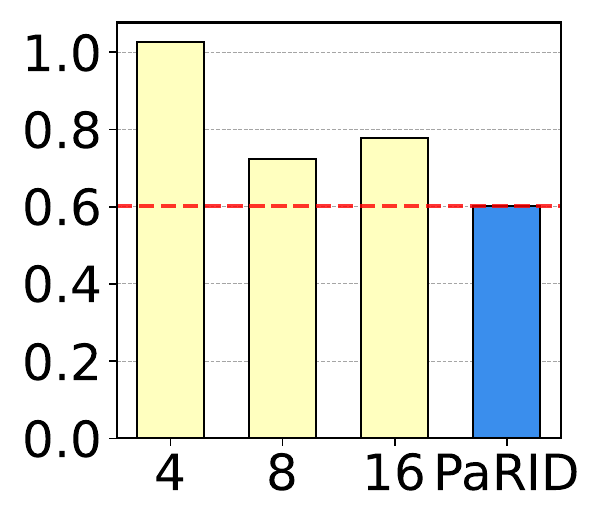}
    }
    \hspace{-3mm}
    \subfigure[LU]{
        \includegraphics[width=0.24\columnwidth]{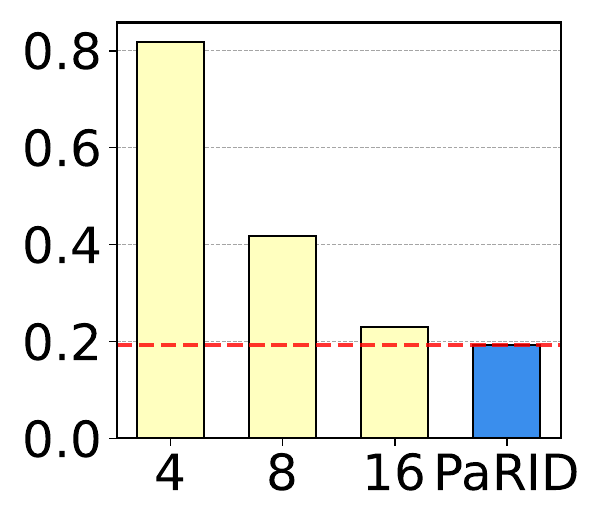}
    }

    \vspace{-3mm}
    \caption{Overall execution time: \tech vs a fixed-thread setting. For each subfigure, x-axis indicates \tech or thread settings, and y-axis is normalized runtime. The red line highlights the normalized runtime of \tech.}
    \label{fig:overall-thread}
\end{figure}

In addition to the baseline method, we also compare \tech with fixed-thread settings.
A fixed-thread setting of $N$ means that all OpenMP kernels in the program are executed with $N$ threads at runtime.
Figure~\ref{fig:overall-thread} presents the results for fixed-thread counts of \{4, 8, 16\}, which consistently achieve top performance across benchmarks, better than \{1, 2, 28, 32\}.
This is also observed in Figure~\ref{fig:thread-sensitivity-study} (Section~\ref{sec:thread-sensitivity-study}).
For IS, EP, CG, MG, and FT, \tech successfully selects the optimal thread configuration.
Specifically, IS and CG contain frequent synchronization, both inter-kernel and intra-kernel, so \tech selects moderate thread counts.
In contrast, EP and FT are embarrassingly parallel and benefit from using more threads without reaching the core limit.
Notably, in BT, SP, and LU, \tech outperforms all fixed-thread settings.
For example, in SP, \tech achieves 20.16\% faster execution than the best fixed setting (8 threads).
These benchmarks represent larger workloads from Computational Fluid Dynamics (CFD) applications, each containing diverse computational patterns~\cite{afzal2017parallelization}.
In BT, for instance, flux difference calculations are embarrassingly parallel, while L2-norm computations involve frequent synchronization~\cite{stone1975parallel}.
\tech's per-kernel tuning enables it to adaptively assign optimal thread counts to these heterogeneous regions, leading to superior performance on large, complex workloads.
Although fixed-thread settings can achieve comparable performance, they are not practical in production environments.
Since real-world applications often run for hours to months~\cite{shaw2021anton}, exhaustively tuning thread counts through trial-and-error methods incurs prohibitive cost.
In contrast, \tech determines these settings at compile time, offering performance portability without any expensive dynamic profiling efforts.

Note that \tech demonstrates high compatibility with OpenMP-based parallel programs, even in the presence of complex inter-thread communication patterns, where none of the existing instruction duplication methods can handle effectively~\cite{huang2022mitigating,he2023demystifying,didehban2016nzdc,didehban2023generic}.
We also evaluate state-of-the-art implementations of serial instruction duplication techniques for comparison.
On average, their normalized execution time is 2.44, significantly higher than \tech's 0.59, highlighting the effectiveness of our parallel-aware design in \ding{202}.

\begin{figure}[h]
\centering
\includegraphics[width=1.0\linewidth]{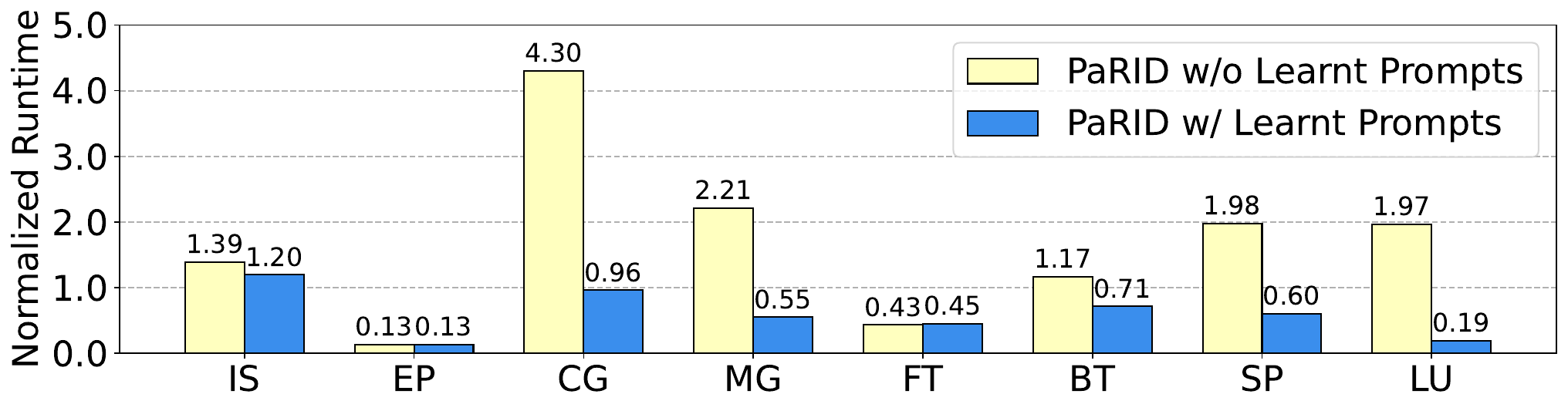}
\vspace{-6mm}
\caption{Impact of \textit{Learned Prompts} for \tech.}
\label{fig:ablation-study}
\end{figure}

\textbf{\textit{Ablation Study for Learned Prompts}}.
We conduct an ablation study by disabling the Learned Prompts (i.e., eight \textit{Findings}) in the \ding{203} stage of \tech workflow.
For each benchmark, we apply \tech with and without Learned Prompts and measure the normalized execution time.
The results are shown in Figure~\ref{fig:ablation-study}.
Without prompt refinement, \tech achieves comparable performance only on IS, EP, and FT.
For the remaining benchmarks, the Learned Prompts significantly enhance the LLM Engine's thread selection, resulting in up to $\sim$9$\times$ speedup (in LU) compared to \tech without prompts.
This performance gap arises because, without these prompts, the LLM tends to make overly simplistic assumptions: assigning 32 threads to kernels without barriers (assuming they are embarrassingly parallel), and 16 threads to those with synchronization.
These default choices often fail to characterize dynamic behaviors introduced by instruction duplication.
This result demonstrates the rationale for the design of the offline Thread Sensitivity Study in \tech.

\begin{table}[h]
\renewcommand{\arraystretch}{0.9}
\centering
\footnotesize
\begin{tabular}{cp{1.1cm}p{1.1cm}p{1.1cm}p{1.1cm}p{1.1cm}} \toprule
\textbf{Name} & \textbf{Class S} & \textbf{Class W} & \textbf{Class A} & \textbf{Class B} & \textbf{Class C} \\\midrule
IS & Keys=16 & Keys=20 & Keys=23 & Keys=25 & Keys=27 \\
EP & M=24 & M=25 & M=28 & M=30 & M=32\\
CG & NA=1400 & NA=7000 & NA=14000 & NA=75000 & NA=150000\\
MG & 32$^3$ & 64$^3$ & 256$^3$ & 256$^3$ & 512$^3$ \\
FT & 64$^3$ & 128$^3$ & 128$\times$256$^2$ & 512$\times$256$^2$ & 512$^3$\\
BT & 12/60 & 24/200 & 64/200 & 102/200 & 162/200\\
SP & 12/100 & 36/400 & 64/400 & 102/400 & 162/400\\
LU & 12/50 & 33/300 & 64/250 & 102/250 & 162/250\\\bottomrule
\end{tabular}
\vspace{3mm}
\caption{Problem size for different NPB input classes~\cite{bailey1991parallel,bailey2010parallel}. X in MG and FT indicates grid size. For A/B in BT, SP, and LU, A and B mean problem size and iteration number.}
\label{tab:input-info}
\end{table}

\textbf{\textit{Input Sensitivity Study}}.
To evaluate \tech's adaptability in production, where a program may be executed with arbitrary inputs, we test \tech on NPB benchmarks using varying input classes.
The NPB suite provides multiple input classes to represent different workload sizes~\cite{bailey2010parallel}.
We focus on five input classes (S, W, A, B, and C) for all NPB benchmarks.
Input details are in Table~\ref{tab:input-info}.
Classes S and W are small workloads, whereas Classes A, B, and C represent standard problem sizes, with each successive class increasing the workload size by $\sim$4$\times$.
For example, Class C in LU corresponds to a $162^3$ grid executed over 250 iterations.
Such numerical details along with explanations are inserted at \textit{Prompt Construction} stage along with \textit{(1) Machine Specification} for guiding LLM inference.

\begin{figure}[h]
    \centering
    \subfigure[IS]{
        \includegraphics[width=0.245\columnwidth]{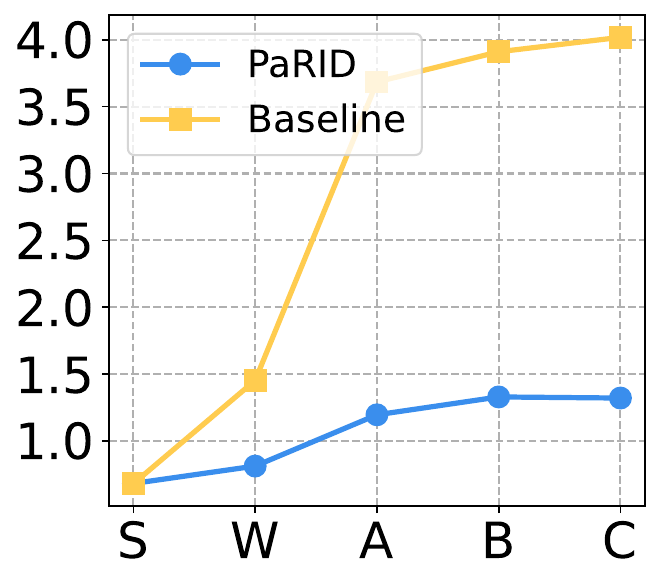}
    }
    \hspace{-3mm}
    \subfigure[EP]{
        \includegraphics[width=0.245\columnwidth]{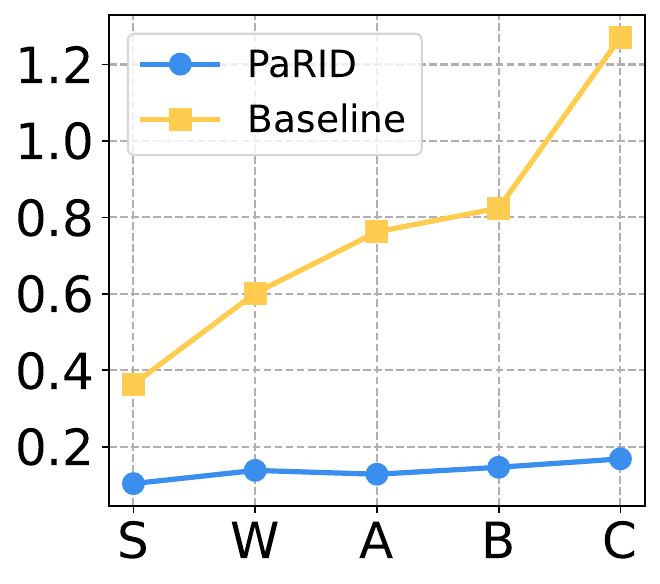}
    }
    \hspace{-3mm}
    \subfigure[CG]{
        \includegraphics[width=0.245\columnwidth]{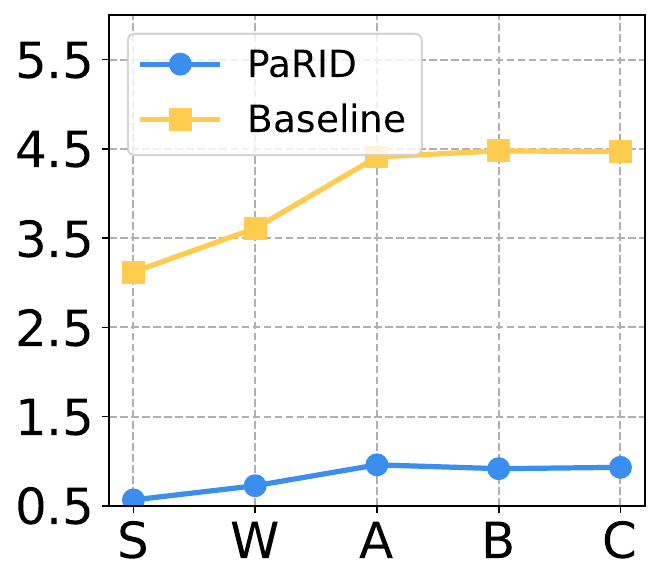}
    }
    \hspace{-3mm}
    \subfigure[MG]{
        \includegraphics[width=0.245\columnwidth]{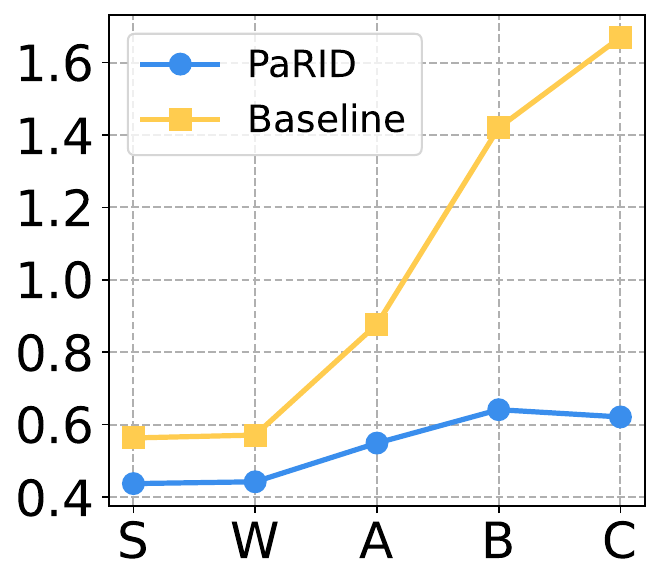}
    }\\\vspace{-2mm}
    
    \subfigure[FT]{
        \includegraphics[width=0.245\columnwidth]{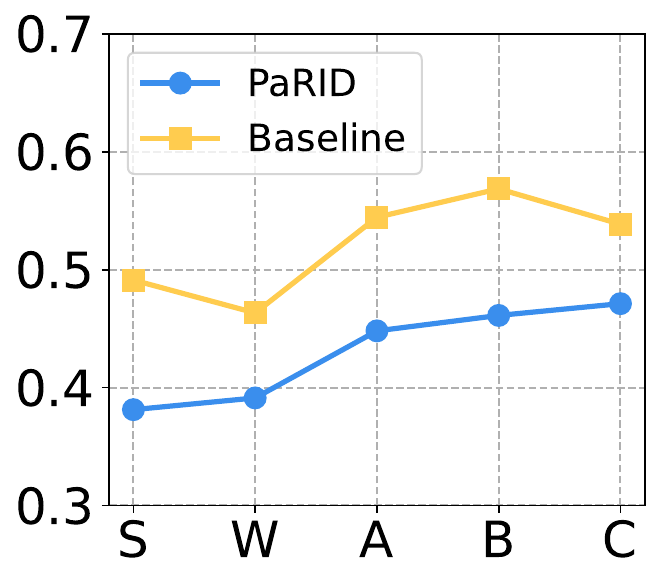}
    }
    \hspace{-3mm}
    \subfigure[BT]{
        \includegraphics[width=0.245\columnwidth]{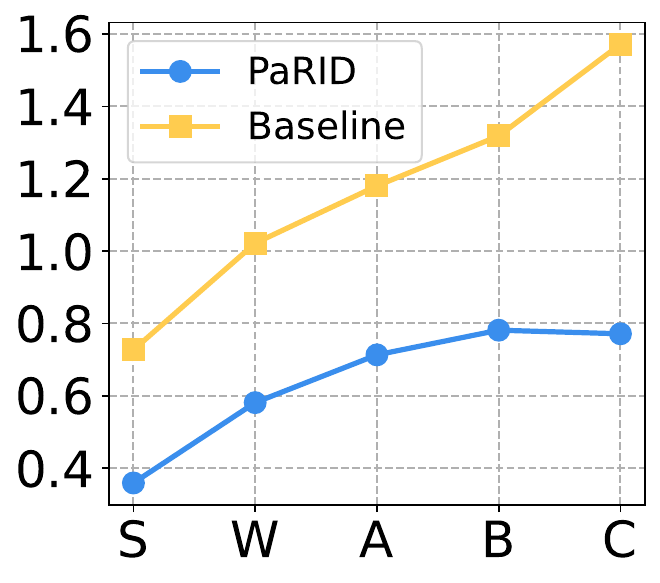}
    }
    \hspace{-3mm}
    \subfigure[SP]{
        \includegraphics[width=0.245\columnwidth]{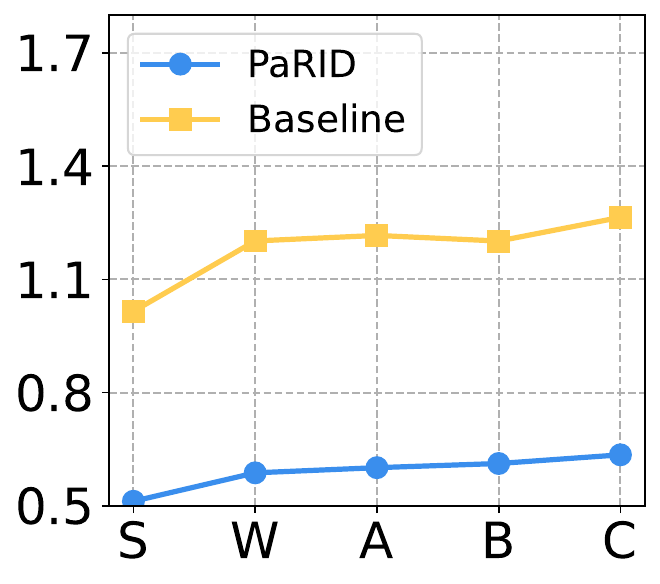}
    }
    \hspace{-3mm}
    \subfigure[LU]{
        \includegraphics[width=0.245\columnwidth]{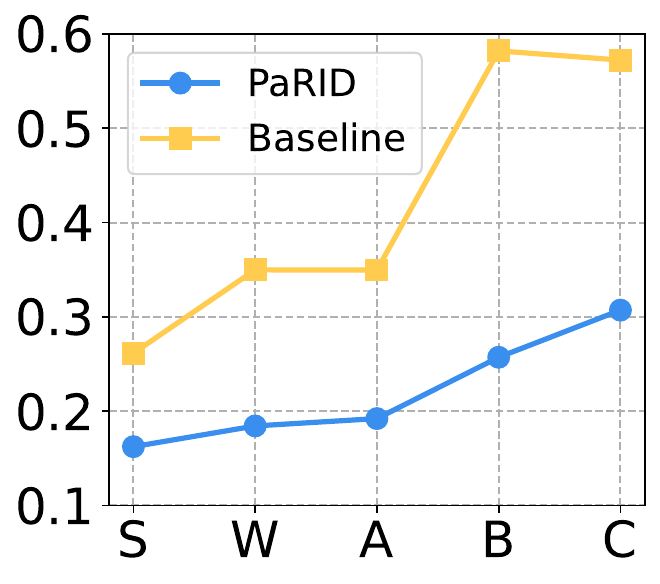}
    }\\
    \vspace{-3mm}
    \caption{\tech with different inputs, where x- and y-axis for each figure mean input classes and normalized runtime.}
    \label{fig:input-sensitivity-study}
\end{figure}

Results are shown in Figure~\ref{fig:input-sensitivity-study}.
Across all input classes, \tech consistently outperforms the baseline method, up to 6.55\% speed up in CG. 
For 3 large CFD workloads, it delivers up to 126.39\% speedup in LU, 103.73\% in BT, and 104.49\% in SP.
This improvement stems from the fact that the baseline routinely uses the maximum number of threads, which becomes increasingly suboptimal as workload sizes grow.
This trend is observed in IS (from Class W to A), EP (from Class B to C), MG (from Class A to B), and LU (from Class A to B).
With larger inputs, the per-thread floating-point workload becomes more intensive, potentially leading to degraded performance due to increased pressure on core resources.
In contrast, \tech dynamically adjusts its thread strategy based on input characteristics.
For instance, in LU, the OpenMP kernel responsible for computing \verb|erhs()| becomes increasingly memory- and computation-intensive for inputs larger than Class B.
\tech identifies this shift and selects a moderate thread count to avoid the performance drop observed in the baseline method between Class A and B.
Similar observations are also observed for fixed-thread settings.

\textbf{\textit{Entire \tech Workflow Time}}.
Executing \tech incurs negligible compile-time overhead, as the only expensive component \textit{Thread Sensitivity Study} is performed offline.
The resulting \textit{Findings} are generalizable and can be reused across different programs at runtime.
The primary runtime cost of \tech lies in LLM inference, which is lightweight: from less than 1 second for IS to $\sim$15 seconds for BT, even without leveraging inference acceleration techniques such as \texttt{\small FlashAttention}.
Importantly, \tech workflow is also a one-time cost.
Once instruction duplication is inserted, \tech enables efficient error detection during parallel execution, demonstrating its practicality for real-world scenarios.

\subsection{Fault Coverage}
\label{sec:fault-coverage}

We evaluate the error detection effectiveness of \tech in this section, using LLFI~\cite{lu2015llfi,agarwal2022lltfi} as our fault injection tool. The choice of LLFI is motivated by two key reasons. First, compared to assembly-level tools, LLFI provides accurate insights into program error propagation behaviors, particularly in terms of SDC rates~\cite{palazzi2019tale}. Second, LLFI (v15~\cite{agarwal2022lltfi}) is fully compatible with our toolset and can be seamlessly extended to support fault injection for OpenMP programs.

To simulate faults in the computational elements of the processor as defined by our fault model (Section~\ref{sec:fault-model}), our fault injection methodology can be explained as follows. 
For each fault injection trial, we randomly sample a site (i.e., a bit in a register involved in execution) and inject a single-bit-flip fault using LLFI~\cite{lu2015llfi,agarwal2022lltfi}.
After program execution is finished, we compare the results between current trial and fault-free execution to verify correctness under PaRID protection.
This approach mirrors methodologies used in prior works~\cite{yang2021enabling,li2018modeling,sangchoolie2017one}. 
For each benchmark, the injection process is repeated 1,000 times, yielding an error bar of <3.1\% for 95\% confidence intervals. 
Fault coverage is assessed by measuring the silent data corruption (SDC) rate, as it is the most severe failure outcome, often bypassing recovery mechanisms~\cite{hursey2007design}.
Our results show that \tech consistently reduces the SDC rate to 0\%, achieving full detection across all benchmarks, regardless of whether they are serial or OpenMP parallel programs or the thread count used.

Parallel-aware Code Transformation (\ding{202}) in \tech also supports a \textit{selective duplication} mechanism~\cite{laguna2016ipas,lu2014sdctune,huang2022mitigating}.
\tech enables both instruction-level and kernel-level selective duplication: the former duplicates only vulnerable instructions, while the latter duplicates instructions only in vulnerable OpenMP kernels.
Unlike serial programs, quantifying the fault coverage and runtime of selective duplication in multithreaded parallel applications requires non-trivial analysis.
We consider this as our future work.


\section{Discussion}


\subsection{Other Hardware Platforms}
In our evaluation (Sections~\ref{sec:thread-sensitivity-study} and~\ref{sec:eva-plantform}), we primarily use an Intel Xeon CPU with 28 cores.
We observe that processor configuration directly impacts the optimal thread count for reducing execution time when applying \tech.
For example, in Section~\ref{sec:new-evaluation}, a thread count of 16 reflects aggressive parallelism on this machine, while 4 and 8 threads represent more moderate configurations.
However, on platforms with significantly more cores (a CPU with stronger parallel capability), these same thread counts become relatively conservative.
To ensure the generalizability of our \textit{Findings}, we avoid specifying exact thread counts in this work.
We further evaluate \tech on a separate system with an AMD EPYC processor (128 cores).
Except for machine specifications, all the rest prompts remain the same.
In this setting, \tech continues to outperform the baseline method, which defaults to using the full core count (128 threads).
For instance, in the EP benchmark, \tech assigns 64 threads to each of the two OpenMP kernels, achieving a 5.37$\times$ speedup (measured by normalized runtime).
These results demonstrate \tech's compatibility across different hardware platforms.

\subsection{Threats to Validity}
One major concern for deploying \tech in production settings is the reliability of the LLM Engine.
Although large language models, especially high-end models such as GPT-4~\cite{achiam2023gpt} and Deepseek R1~\cite{guo2025deepseek}, demonstrate strong inference capabilities, the accuracy of smaller device-scale models remains an open question.
For example, LLMs with similar parameter sizes to the \textit{Phi-3.5-mini-instruct} used in this work have unknown capabilities in predicting dynamic program features.
To investigate this, we replace the original LLM in \tech with \textit{deepseek-coder-7b-instruct-v1.5}~\cite{guo2024deepseek}, a 6.91B code-focused LLM that is publicly available on Huggingface.
We run local inference using the same prompt structure defined in \tech.
Initially, the model outputs a Python script for predicting performance via linear regression, rather than directly responding with thread recommendations.
However, after refining the phrasing in the natural language portion of the prompts, while keeping all eight \textit{Findings} unchanged, the model produces thread decisions consistent with those reported in Section~\ref{sec:new-evaluation}.
This result suggests that the LLM engine in \tech is replaceable but requires tuning of prompt languages.
It is also worth noting that even state of the art LLMs, due to limited domain knowledge and training data, cannot infer instruction duplication overhead.
The eight \textit{Findings} remain essential for steering the model toward accurate predictions.

\subsection{Extending \tech Protection to Other ISAs}
In this work, we perform \tech code transformation and fault injection at the LLVM IR level. While this is well-justified and supported by existing works~\cite{huang2022mitigating,kalra2020armorall,laguna2016ipas,li2018modeling,anwer2020gpu}, extending \tech to other instruction set architectures (ISAs) remains a valuable direction for future exploration.
Recall from Section~\ref{sec:high-level-ompid}, the key components of \tech~\ding{202} include \textit{(1) Identifying Parallel Regions}, \textit{(2) Annotating Eligible Instructions}, and \textit{(3) Dataflow Analysis for Parallel}. When applying \tech to other instruction sets, such as the x86 ISA, the transformation process remains largely consistent, with some slight differences.
Specifically, while components (2) and (3) are unaffected, the outermost OpenMP directive is not explicitly translated into a function call in the x86 ISA. Instead, OpenMP kernels execute immediately following an OpenMP runtime function, {\small\texttt{kmpc\_fork\_call}}, which serves as a clear marker for identifying parallel regions. 
This ensures that OpenMP subroutines can still be recognized as distinct entities in the assembly code.
Moreover, due to the flexibility of LLVM design, IR code with \tech protection can be compiled naturally into other ISA formats.
In a nutshell, while \tech is implemented at the LLVM IR level, its design is generic and not restricted to LLVM.
\section{Related Works}

\subsection{Instruction Duplication}
Instruction duplication has been extensively explored over the past two decades~\cite{oh2002error,reis2005swift,lu2014sdctune,kalra2020armorall,didehban2018compiler,mahmoud2018optimizing}.
This technique operates independently of the program’s algorithm, making it entirely program-agnostic.
Reis et al. introduced SWIFT~\cite{reis2005swift} and SWIFT-R~\cite{reis2007automatic}, which leverage instruction-level redundancy to detect errors and recover correct program execution.
Laguna et al.~\cite{laguna2016ipas} employed a machine learning model to predict instruction vulnerability, enabling selective instruction duplication to achieve high fault coverage and low runtime overhead.
This work also discussed instruction duplication on process-level parallelism (with MPI).
Didehban et al.~\cite{didehban2016nzdc} duplicated instructions at the microarchitecture-level for reducing SDC in ARM CPUs.
Huang et al.~\cite{huang2023characterizing} characterized the runtime overhead variations of instruction duplication from both compiler-level and microarchitecture-level perspectives.
Although effective, their approach targets only serial programs and neglects addressing multithreading parallel programs, which is crucial in production.

\subsection{Error Resilience on Parallel Architectures}
Reliability research has also focused on exploring error propagation and redundancy-based detection in parallel architectures~\cite{kuvaiskii2016elzar,kalra2020armorall,mahmoud2018optimizing,yim2011hauberk,yang2021enabling,yang2021sugar}.
Yim et al.~\cite{yim2011hauberk} incorporated duplication along with checksums and accumulation-based range checking for lightweight protection for GPGPU programs.
Kuvaiskii et al.~\cite{kuvaiskii2016elzar} adopted AVX, a SIMD instruction set for Intel CPU, to conduct fast error detection and execution recovery.
Mahmoud et al.~\cite{mahmoud2018optimizing} performed code transformations at the SASS level to enable efficient and effective error detection on NVIDIA GPUs.
Yang et al.~\cite{yang2021enabling} selectively copies unreliable threads for efficient error resilience for GPGPU applications.
Fang et al.~\cite{fang2014evaluating} extended LLFI to support OpenMP parallel programs but focused solely on studying error propagation behaviors rather than implementing an effective error detection mechanism.
Wang et al.~\cite{wang2007compiler} leveraged thread-level redundancy for soft error detection.
Compared with existing works, \tech not only supports multithreaded parallel programming models but also ensures competitive execution time through LLM-inferred thread tuning, making it better suited for production environments.

\section{Conclusion}

In this work, we first conduct an initial study to understand how parallel programs are reflected at the instruction level through static analysis, and to assess how effectively device-scale LLMs can capture dynamic program features through dynamic analysis.
Based on these insights, we propose \tech, an LLM-tuned code transformation framework that achieves full compatibility with OpenMP-based parallel programs while providing optimal thread tuning through LLM inference.
We also introduce a Thread Sensitivity Study and summarize eight generalizable \textit{Findings}.
Our evaluation demonstrates that \tech (1) maintains compatibility with OpenMP programs, (2) efficiently reduces execution time through kernel-specific thread tuning, (3) consistently adapts to varying inputs, and (4) preserves full soft error detection effectiveness.

\bibliographystyle{ACM-Reference-Format}
\bibliography{reference}

\end{document}